\documentclass[10pt,journal,compsoc]{IEEEtran}
\usepackage{cite}
\usepackage{amsmath,amssymb,amsfonts}
\usepackage{subfigure}
\usepackage[linesnumbered,ruled,vlined]{algorithm2e}
\usepackage{algorithmic}
\usepackage{graphicx}
\usepackage{textcomp}
\usepackage{xcolor}
\usepackage{multirow}
\usepackage{array}
\usepackage{bm}
\newcommand{\badown}[1]{\scriptsize{\textcolor{myred}{$\downarrow$ #1}}}
\newcommand{\baup}[1]{\scriptsize{\textcolor{mygreen}{$\uparrow$ #1}}}
\definecolor{mygreen}{RGB}{0 139 139}
\definecolor{myred}{RGB}{238 99 99}

\def\BibTeX{{\rm B\kern-.05em{\sc i\kern-.025em b}\kern-.08em
    T\kern-.1667em\lower.7ex\hbox{E}\kern-.125emX}}
\begin{document}

\title{
Preventing Non-intrusive Load Monitoring Privacy Invasion: A Precise Adversarial Attack Scheme for Networked Smart Meters
}
\author{Jialing He,
        Jiacheng Wang,
        Ning Wang,
        Shangwei Guo,
        Liehuang Zhu,~\IEEEmembership{Senior Member,~IEEE,}
        Dusit Niyato~\IEEEmembership{Fellow,~IEEE}
        and Tao Xiang,~\IEEEmembership{Senior Member,~IEEE}
        \thanks{Jialing He, Ning Wang, Shangwei Guo, and Tao Xiang are with the College of Computer Science, Chongqing University, Chongqing, China.}
        \thanks{Jiacheng Wang and Dusit Niyato are with Nanyang Technological University, Singapore.}
        \thanks{Liehuang Zhu is with School of Cyberspace Science and Technology, Beijing Institute of Technology, China.}
        \thanks {Corresponding authors: Jiacheng Wang (jiacheng.wang@ntu.edu.sg) and Ning Wang (nwang5@cqu.edu.cn).}
        }

\markboth{JOURNAL OF LATEX CLASS FILES,VOL. XX, NO. XX,~2024}
{He \MakeLowercase{\textit{et al.}}: Bare Demo of IEEEtran.cls for IEEE Journals}

\IEEEtitleabstractindextext{

\begin{abstract}
Smart grid, through networked smart meters employing the non-intrusive load monitoring (NILM) technique, can considerably discern the usage patterns of residential appliances. However, this technique also incurs privacy leakage. To address this issue, we propose an innovative scheme based on adversarial attack in this paper. The scheme effectively prevents NILM models from violating appliance-level privacy, while also ensuring accurate billing calculation for users. To achieve this objective, we overcome two primary challenges. First, as NILM models fall under the category of time-series regression models, direct application of traditional adversarial attacks designed for classification tasks is not feasible. To tackle this issue, we formulate a novel adversarial attack problem tailored specifically for NILM and providing a theoretical foundation for utilizing the Jacobian of the NILM model to generate imperceptible perturbations. Leveraging the Jacobian, our scheme can produce perturbations, which effectively misleads the signal prediction of NILM models to safeguard users' appliance-level privacy.
The second challenge pertains to fundamental utility requirements, where existing adversarial attack schemes struggle to achieve accurate billing calculation for users. To handle this problem, we introduce an additional constraint, mandating that the sum of added perturbations within a billing period must be precisely zero.
Experimental validation on real-world power datasets REDD and UK-DALE demonstrates the efficacy of our proposed solutions, which can significantly amplify the discrepancy between the output of the targeted NILM model and the actual power signal of appliances, and enable accurate billing at the same time. Additionally, our solutions exhibit transferability, making the generated perturbation signal from one target model applicable to other diverse NILM models.
\end{abstract}

}
\maketitle

\IEEEdisplaynontitleabstractindextext

\IEEEpeerreviewmaketitle

\section{Introduction}
The smart grid, a prominent realization of advanced Internet of Things (IoT) systems, has significantly improved power efficiency through the integration of networked smart meters~\cite{yang2022energy}. These smart meters, coupled with mobile technologies, enable real-time monitoring of household power consumption data, transmitted frequently to cloud servers for centralized processing~\cite{ke2018efficiency}. This integration allows energy providers to accurately forecast future energy demands, ultimately enhancing the efficiency of the smart grid. However, this convenience comes at the cost of user privacy~\cite{kumar2019smart,mccary2015smart}. For example, adversaries who eavesdrop on the transmission of power consumption data from mobile-connected smart meters to utility providers or exploit data made accessible to unauthorized personnel within utility companies could leverage advanced algorithms to infer detailed user behaviors and activities.

Non-Intrusive Load Monitoring (NILM) is a pivotal technique utilized to infer power consumption patterns~\cite{hart1992nonintrusive}. This approach leverages statistical models, such as Hidden Markov Models (HMMs)~\cite{kolter2012approximate,guo2014home,kong2016extensible}, to determine the on-off states of appliances from the metered data. Recent studies indicate that it can also deploy Deep Neural Networks (DNNs) to disaggregate the aggregate power signal recorded by the smart meter into the power consumption data of individual appliances~\cite{kelly2015neural,zhang2018sequence,shin2019subtask,he2023infocus,yue2020bert4nilm,He2023msdc,chen2019scale,piccialli2021improving}. This process, known as energy disaggregation, enables the inference of users' appliance preferences, daily routines, and even their home occupancy~\cite{jin2017virtual}. The acquisition of such sensitive information by adversaries could be manipulated for intrusive marketing strategies or, in extreme scenarios, could pose a threat to users' personal safety~\cite{kumar2019smart,philip2021internet}.
%

As smart meters play an increasingly central role in modern power grid systems, their reliance on mobile networks for the transmission of real-time data makes safeguarding user privacy an urgent priority. Current privacy-preserving schemes, such as Differential Privacy (DP)~\cite{lou2017cost,hassan2022differentially,lyu2018ppfa,won2015privacy,zheng2021decentralized,wang2020privacy}, attempt to obfuscate power consumption data by injecting noise.
The inherent characteristics of DP, which inject noise into statistical outputs derived from a fixed dataset, result in a constant sensitivity. Consequently, the noise, when scaled by this fixed sensitivity, may not be well-suited for power consumption data that varies over time. Additionally, many DP approaches necessitate the involvement of a third party to compute the sensitivity, potentially introducing privacy concerns.

Adversarial attacks, initially devised to produce imperceptible perturbations inducing misclassifications in targeted DNNs, provide a promising avenue for misleading NILM models.
Fawaz \emph{et al.}\cite{fawaz2019adversarial} were the first to adapt existing attacks designed for image classifiers~\cite{goodfellow2014explaining,Kurakin2017,madry2017towards,dong2018boosting,wu2018understanding} to deceive residual network-based time-series classification models. Karim \textmd{et al.}\cite{karim2020adversarial} proposed the use of an adversarial transformation network on a distilled model, enabling attacks on various time-series classification models. Pialla \emph{et al.}~\cite{pialla2022smooth} introduced the Smooth Gradient Method (SGM) to generate smoother perturbations by imposing a smoothness condition. Additionally, several adversarial attacks in the smart grid domain have been proposed, targeting tasks such as power quality disturbance (PQD) classification~\cite{tian2021adversarial,khan2024adaptedge}, power system event classification~\cite{cheng2022adversarial}, and electricity theft detection~\cite{takiddin2022robust}. Notably, all these tasks involve time-series classification.

Subsequently, some studies extended adversarial attacks to time-series regression models. Tang \emph{et al.}~\cite{wu2022small} and Wu \emph{et al.}~\cite{tang2021adversarial} followed the basic approach of attacking classification models and provided gradient-based methods for attacking time-series prediction models. Later, Gupta \emph{et al.}~\cite{gupta2021adversarial} formally formulated the adversarial problem for regression tasks and devised an adversarial attacker that exploits the algebraic properties of the Jacobian. Kong et al.~\cite{kong2023adversarial} further optimized the study of Gupta \emph{et al.}~\cite{gupta2021adversarial} by redefining the attack objective in alignment with the characteristics of regression problems. However, the approach proposed by Gupta \emph{et al.}~\cite{gupta2021adversarial} does not incorporate constraints on the attack objective that can be directly applied for deceiving NILM models. Moreover, the method by Kong et al.~\cite{kong2023adversarial} is only applicable for regressors with a one-time output, whereas an NILM model is a time-series output regressor. Additionally, none of these methods incorporate constraints on adding perturbations that can satisfy utility requirements

This study aims to bridge the gap within the field through the introduction of a pioneering scheme, Adversarial Non-intrusive Load Monitoring (ADV-NILM). ADV-NILM entails the proposition and formulation of the adversarial attack problem for NILM models. The core algorithm of ADV-NILM is architectured through an extensive theoretical analysis rooted in the algebraic properties of the Jacobian intrinsic to the NILM model. By leveraging this algorithm, ADV-NILM is capable of generating an optimal, imperceptible perturbation signal that can potentially impair the performance of the targeted NILM models. This, in effect, serves to protect users' privacy at the appliance level. To further ensure accurate billing calculations for users, the adversarial attack problem is augmented by integrating an additional constraint for the perturbations. Our primary contributions are summarized as follows:

\begin{itemize}
  \item We delineate the adversarial attack problem for NILM models and conduct a comprehensive theoretical analysis of the solution. Drawing from this theoretical framework, we propose a solution algorithm that employs the Jacobian information of the target NILM model to generate manipulated signals. These perturbed power measurements successfully induce erroneous outputs from the NILM model.

  \item  To ensure precise billing calculations for users, we further refine and adapt the adversarial attack problem to a practical version. This refinement involves the integration of an additional constraint, mandating that the total sum of added perturbations within a billing period must be zero. To comply with this practical requirement, we modify the proposed solution algorithm accordingly.

   \item Our proposed methodologies are rigorously evaluated through experiments on real-world power datasets, including REDD~\cite{kolter2011redd} and UK-DALE~\cite{kelly2015uk}. The experimental results demonstrate the effectiveness of our approaches in generating perturbations that significantly increase the divergence between the output of the targeted NILM model and the actual power signal of appliances. Notably, the perturbation signal generated for a specific NILM model exhibits transferability and can effectively deceive other NILM models as well.
\end{itemize}
\section{Related Work}
\subsection{DNN-based NILM models}
Since the seminal work of NILM~\cite{hart1992nonintrusive}, numerous solutions have been proposed, primarily centered around statistical inference models like Hidden Markov Models (HMMs). However, these HMM-based approaches make the assumption that each observation (i.e., per power data) is independent of others, which is not reflective of real-life scenarios and hampers practicality. To overcome this limitation, the emergence of deep learning techniques has opened up new avenues for addressing NILM challenges. Leveraging the remarkable learning capabilities of neural networks, the NILM problem can be tackled by training a DNN to map the aggregated power signal to individual appliance power signals.
The pioneering work of Kelly and Knottenbelt~\cite{kelly2015neural} introduced the Neural NILM architecture, incorporating Long Short-Term Memory (LSTM), CNN, and denoising autoencoders. This approach achieved significant advancements over previous statistical inference models. Subsequently, a series of DNN-based schemes aimed at further enhancing performance have been proposed. Notably, Zhang \emph{et al.}~\cite{zhang2018sequence} presented the Seq2seq architecture, a landmark model constructed using convolutional layers and dense layers. This simple yet effective architecture was able to extract latent features, including power thresholds and state change points, leading to substantial improvements.
Building on Seq2seq, subsequent works employed various strategies to boost performance. Some of these strategies include adopting a sequence-to-point output strategy (i.e., outputting the middle point of a sequence instead of the entire sequence)~\cite{zhang2018sequence}, extracting appliance state information as auxiliary data~\cite{shin2019subtask,he2023infocus,He2023msdc}, constructing hybrid models with multiple structures (such as CNN, LSTM, and GAN)~\cite{chen2019scale,piccialli2021improving}, and incorporating attention mechanisms\cite{yue2020bert4nilm}. These endeavors have further pushed the boundaries of NILM model performance.
\subsection{Adversarial attacks}
Adversarial attacks initially gained traction in the realm of image classification tasks, with the objective of altering model outputs by adding imperceptible noise. One of the seminal works in this area was the FGSM introduced by Goodfellow \emph{et al.}~\cite{goodfellow2014explaining}. This single-step attack utilized gradient information from the target DNN to create perturbations. Building upon the success of FGSM, Kurakin \emph{et al.}~\cite{Kurakin2017} and Madry \emph{et al.}~\cite{madry2017towards} proposed iterative versions of the attack, known as the Basic BIM and PGD, respectively. These advances subsequently inspired similar attacks like M-IGSM\cite{dong2018boosting} and vr-IGSM~\cite{wu2018understanding}.
Despite the efficacy of adversarial attacks in image classification tasks, their direct application to time-series models proved challenging or ineffective. Unlike images, modifying a single data point in a time-series may result in noticeable changes. Fawaz \emph{et al.}~\cite{fawaz2019adversarial} were the first to adapt existing attacks designed for image classifiers to deceive residual network-based time-series classification models. Karim \emph{et al.}~\cite{karim2020adversarial} proposed an adversarial transformation network on a distilled model, enabling attacks on various time-series classification models. Pialla \emph{et al.}~\cite{pialla2022smooth} introduced the SGM to generate smoother perturbations by imposing a smoothness condition.
Subsequently, some studies extended adversarial attacks to time-series regression models. Tang et al.\cite{wu2022small} and Wu \emph{et al.}~\cite{tang2021adversarial} followed the basic approach of attacking classification models and provided gradient-based methods for attacking time-series prediction models. However, these prior adversarial attacks were constructed based on the gradient information of target models and lacked a formal theoretical justification.
Addressing this limitation, Gupta \emph{et al.}~\cite{gupta2021adversarial} formulated the adversarial problem for regression tasks and developed an adversarial attacker that leverages the algebraic properties of the Jacobian. Their study provided a theoretical analysis considering multiple perturbation constraints. However, a perturbation constraint specifically suitable for deceiving NILM models was not included in their work. Moreover, the fundamental utility requirement, particularly accurate billing calculation for users, was not addressed in their approaches.

\section{Preliminaries}\label{sec:preliminary}
\subsection{NILM Problem}
The NILM problem aims to deduce the power consumption data of individual appliances from an aggregated power signal. Let $X = \{ x_1, x_2, \ldots , x_T\} \in \mathbb{R}+^T$ denote the aggregated power signal, where $T \in \mathbb{N}+$ represents the total measured time. The investigation considers $N$ appliances, and the power consumption of the $i$th appliance is denoted as $Y^i = \{y_1^i, y_2^i, \ldots, y_T^i\}\in \mathbb{R}_+^T$. A fundamental assumption in the NILM problem is that for all $i \in [N]$ and $t \in [T],$ where $[T]=\{1, 2, \ldots, T\}$ and $[N]=\{1, 2, \ldots, N\}$, we have:
  $x_t=\sum_{i=1}^N y_t^i+u_t+\nu_t$.
$u_t$ represents the total power consumption data of all the other appliances that are out of the investigation's scope and $\nu_t$ is the noise signal.
\subsection{Solution Model for NILM Problem}
Given a set of instances of energy consumption data $\{X, Y^1, Y^2, \ldots, Y^N\}$, solution models for the NILM problem is to infer $y_t^i$ for all $i \in [N]$ and $t \in [T]$, and further generalize to unseen test power signal $X'$.

For the original power consumption data, a very long power sequence, solution models employ the \emph{sliding window} method to operate. The original aggregated power signal is divided into several overlapping $w$-length segments, each of which is one input window for the solution model. Suppose $X_{t,w}=(x_{t-\lfloor\frac{w}{2}\rfloor}, \ldots, x_{t+\lceil\frac{w}{2}\rceil-1})$ denote the input window and $f^i_\theta\colon\mathbb{R}_+^w \rightarrow \mathbb{R}_+^w$ ($\theta$ denotes the parameters of the model and is omitted for simplicity) represents the solution model for the $i$th ($i\in [N]$) appliance, such that:
\begin{equation}\label{eq:solution}
  \hat{Y}^i_{t,w} = f^i(X_{t,w}),
\end{equation}
where $\hat{Y}^i_{t,w}=(\hat{y}_{t-\lfloor\frac{w}{2}\rfloor}, \ldots, \hat{y}_{t+\lceil\frac{w}{2}\rceil-1})$ is the output window, i.e., predicted power consumption for the $i$th appliance.
A solution model for NILM problem can take the form of a statistic model such as HMMs \cite{zia2011hidden, mauch2016novel} or deep neural networks including CNNs, RNNs, and attention mechanisms \cite{zhang2018sequence, d2019transfer, ccimen2020microgrid, shin2019subtask, piccialli2021improving,he2023infocus,yue2020bert4nilm,He2023msdc}. For simplicity, we refer to the solution model for NILM problem as the NILM model in the subsequent section.
\section{Threat model}\label{sec:threat}
As previously noted, the NILM model has the ability to deduce the power signal of installed appliances from the aggregated power reading recorded, thereby inferring the user's power consumption pattern.
In this research, the objective for users is to generate a perturbation signal which, when superimposed on the original aggregated signal, can deceive the target NILM model into producing predictions that significantly deviate from the true appliances' signal. As a result, the actual power consumption patterns of users at the appliance-level can be effectively concealed. Since this method follows the adversarial attack paradigm, we refer to the user objective as the ``attack objective" and the users' capabilities as the ``attack capabilities."

\noindent\textbf{Attack Objective. }
First, the primary attack objective is to deceive the target NILM model into producing appliance-level signals that significantly deviate from the true values.
Furthermore, based on this attack objective, users are required to maintain the fundamental billing operation for the smart grid. This stipulates that the real-time added perturbation signal should not affect the accuracy of the cumulative sum of power within the billing period.

\noindent\textbf{Attack Capabilities. }
Both white-box~\cite{goodfellow2014explaining} and black-box~\cite{papernot2017practical} attack settings are considered in this study. In the white-box setting, it is assumed that the user has access to the target NILM model $f(\cdot)$, including its architecture and parameters $\theta$. Consequently, the user can utilize the target model's parameters to generate the perturbation signal. Conversely, in the black-box setting, the user lacks access to the target NILM model. However, it is assumed that the user can train an alternative NILM model $f'(\cdot)$, referred to as the shadow model, whose architecture may differ from $f(\cdot)$. The user can subsequently generate perturbations using the trained model $f'(\cdot)$. For both black-box and white-box settings, the allowable magnitude of the user-added perturbation is limited.
\section{Problem formulation}\label{sec:problem}
In the preceding section, we have intuitively outlined the attack objective for users.
In this section, we formally define the problem to realize this objective, terming it the \emph{adversarial NILM problem}. Subsequently, we extend this formulation to include the \emph{practical adversarial NILM problem}, which ensures the accurate billing computation for users.
\begin{figure*}[htp]
	\centering
	\includegraphics[width=1.0\textwidth]{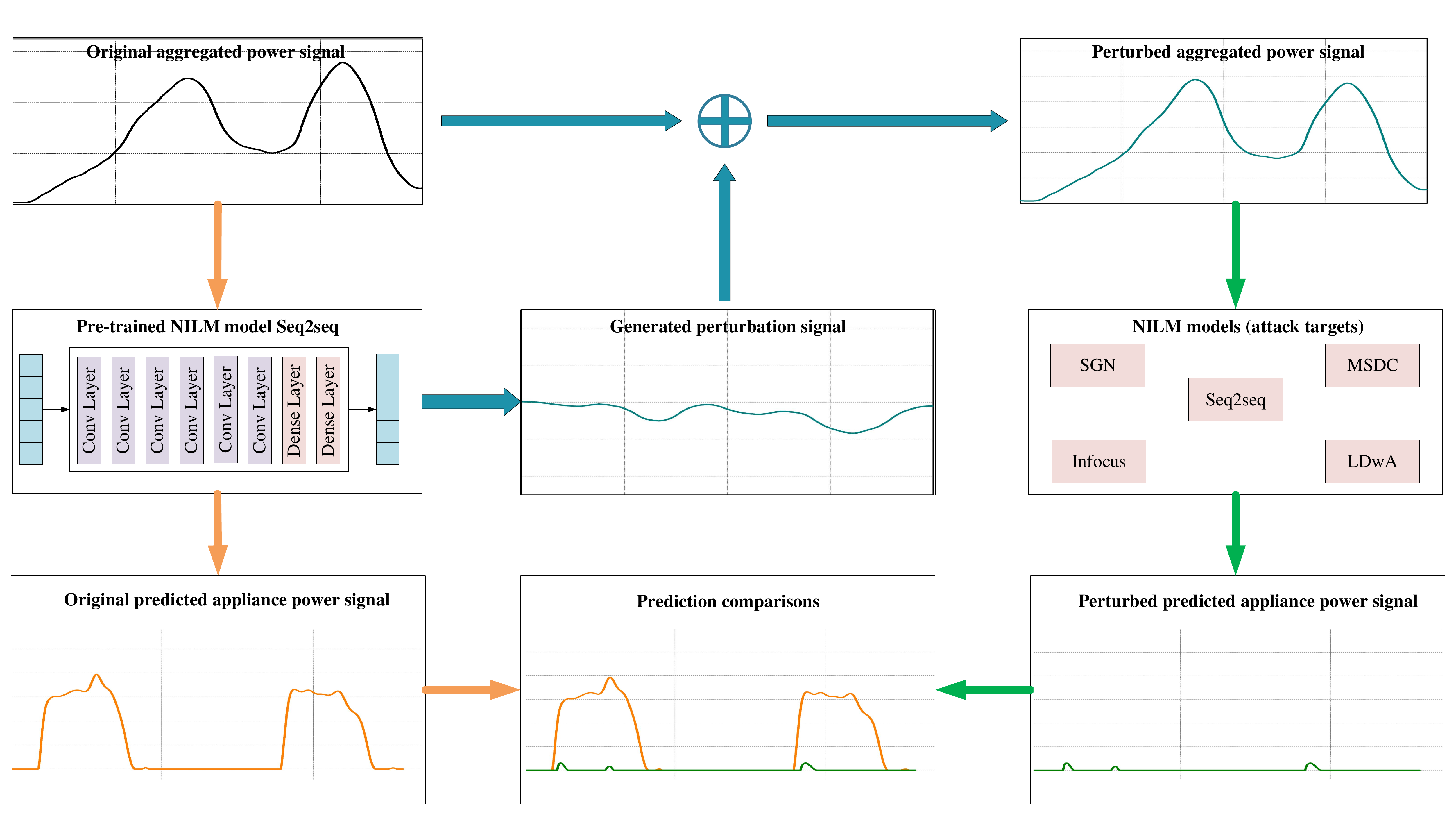} 
	\caption{Framework of our ADV-NILM. ADV-NILM generates perturbation signals using a trained NILM model, which are subsequently added to the original smart meter readings. The perturbed readings effectively deceive adversaries, causing mainstream NILM models to infer appliance-level power usage that significantly deviates from the actual values.}
	\label{fig:framework}
\end{figure*}
\subsection{Adversarial NILM Problem}
The \emph{adversarial NILM problem} aims to generate a perturbed aggregated power signal that deceives the NILM solution model into producing a power signal that significantly deviates from the actual data of the target appliances. More formally, given an aggregated power signal $X_{t,w} = (x_{t-\lfloor\frac{w}{2}\rfloor}, \ldots, x_{t+\lceil\frac{w}{2}\rceil-1})$, the problem is, for $\forall i\in [N]$, finding the ``worst perturbation" $\tilde{E}^i_{t,w}=(\tilde{e}^i_{t-\lfloor\frac{w}{2}\rfloor}, \ldots, \tilde{e}^i_{t+\lceil\frac{w}{2}\rceil-1}) \in \mathbb{R}^w$ such that
\begin{equation}\label{eq:adv_NILM}
  \tilde{E}^i_{t,w} = \mathop{\arg\max}\limits_{E^i_{t,w}\in C^i_\delta}\parallel f^i(X_{t,w}+E^i_{t,w})-Y_{t,w}^i\parallel_1,
\end{equation}
where $C^i_\delta$ is a set that limits the amount of perturbation added and is defined as
\begin{equation}\label{eq:constraint}
  C^i_\delta = \{E^i_{t,w} \in \mathbb{R}^w | \parallel E^i_{t,w} \parallel_{\infty}\leq \delta\}.
\end{equation}
Here we use $Y^i_{t,w}=(y^i_{t-\lfloor\frac{w}{2}\rfloor}, \ldots, y^i_{t+\lceil\frac{w}{2}\rceil-1})$ to denote the truth power consumption data for the $i$th appliance, $\tilde{X}_{t,w}=X_{t,w}+E_{t,w}= (\tilde{x}_{t-\lfloor\frac{w}{2}\rfloor}, \ldots, \tilde{x}_{t+\lceil\frac{w}{2}\rceil-1})$ represents the generated adversarial power signal, $\parallel\cdot\parallel_{\infty}$ denotes the matrix norm, and $\delta$ corresponds to the maximum allowed perturbation.
\subsection{Practical Adversarial NILM Problem}
Eq.~(\ref{eq:constraint}) in the previous \emph{adversarial NILM problem} limits the magnitude of the perturbation in real-time, specifically per time step.  However, this constraint alone falls short of meeting the practical requirement. In real-world scenarios, it is crucial that the additional perturbation does not impact the billing operation. To ensure this, we propose the following constraint:
\begin{equation}\label{eq:pra}
    C^i_\delta = \{E^i_{t,w} \in \mathbb{R}^w | \sum E^i_{t,w} = 0\},
\end{equation}
 where $\sum E_{t,w}^i = e^i_{t-\lfloor\frac{w}{2}\rfloor}+ \cdots +e^i_{t+\lceil\frac{w}{2}\rceil-1}$. It ensures that the sum of consecutive perturbations added within a window is zero. This guarantees that the accumulated perturbed power signal remains the same as the original signal, thereby preventing errors in the electricity bill calculation.

We are now ready to present the \emph{practical adversarial NILM problem}. Formally, given an aggregated power signal $X_{t,w} = (x_{t-\lfloor\frac{w}{2}\rfloor}, \ldots, x_{t+\lceil\frac{w}{2}\rceil-1})$, the problem is, for $\forall i\in [N]$, finding the perturbation $\tilde{E}^i_{t,w}=(\tilde{e}^i_{t-\lfloor\frac{w}{2}\rfloor}, \ldots, \tilde{e}^i_{t+\lceil\frac{w}{2}\rceil-1}) \in \mathbb{R}^w$:
\begin{equation}\label{eq:adv_prac}
\begin{split}
  \tilde{E}^i_{t,w} = \mathop{\arg\max}\limits_{E^i_{t,w}\in C^i_\delta}\parallel f^i(X_{t,w}+E^i_{t,w})-Y_{t,w}^i\parallel_1\\
  s.t.   C^i_\delta = \{E^i_{t,w} \in \mathbb{R}^w | \parallel E^i_{t,w} \parallel_{\infty}\leq \delta \wedge \sum E^i_{t,w} = 0\}&
\end{split}
\end{equation}
\section{Adversarial-NILM Solution Model}\label{sec:model}
The previous section provided a clear problem formulation of the adversarial NILM problem.
In this section, we present our novel scheme, named ADV-NILM, designed to address the adversarial NILM problems.
We provide an overview of the framework, followed by a detailed explanation of its individual components.
\subsection{Overview}
As Fig.~\ref{fig:framework} shows, the ADV-NILM seeks to produce well-crafted perturbed power signal to fool all the NILM models. More specifically, given a window of aggregated power signal, the NILM model can accurately dissect this original aggregated signal into per-appliance signals. ADV-NILM can produce perturbations to perturb the original aggregated signal and deceive the NILM models. The generated perturbations should be as small as possible (Eq.~(\ref{eq:constraint}) gives the constraint for the perturbation) and the difference between the original output and perturbed output of NILM models should be as big as possible (Eq.~(\ref{eq:adv_NILM}) describes this problem). Details of the NILM models and the perturbed-signal generation process follow shortly in this section.
\subsection{NILM Model Description}\label{sec:model-des}
DNN-based NILM models have demonstrated remarkable capabilities in extracting per-appliance power signals, owing to the high expressibility and learnability of neural networks.
The pioneering work of Kelly and Knottenbelt \cite{kelly2015neural} introduced NeuralNILM, which served as the first deep neural network-based architecture for addressing the NILM problem. By integrating networks such as CNN, LSTM, and denoising autoencoders, NeuralNILM outperformed previous statistical and machine learning models, establishing a new benchmark in the field.
Building upon this success, Zhang et al.~\cite{zhang2018sequence} presented Seq2seq, a critical CNN-based NILM model that exhibited considerable performance improvements, subsequently inspiring numerous follow-up solutions. Given the reliability and effectiveness of Seq2seq, we have selected it as our target NILM model for generating perturbations in this study.

Seq2seq employs one CNN $f^i_{power}\colon\mathbb{R}_+^w \rightarrow \mathbb{R}_+^w$ to learn the mapping $\hat{P}_{t,w} = f^i_{power}(X_{t,w})$. More details, please refer to~\cite{zhang2018sequence}. The objective loss function to train Seq2seq is:
\begin{equation}\label{eq:loss}
\arg\min_{\substack{\theta}} \frac{1}{T}\sum_{t=1}^{T}(f^i_{\theta}(X_{t,w}) - Y^i_{t,w})^2,
\end{equation}
where $Y^i_{t,w}$ is the truth power signal of the $i$th appliance.
Seq2seq is trained using the Stochastic Gradient Descent (SGD) optimization algorithm, which utilizes gradients to continuously update the model parameters for minimizing the loss function.

\subsection{Perturbed Signal Generation}
Gupta \emph{et al.}~\cite{gupta2021adversarial} proposed an adversarial attacker for regression problems. However, the proposed attack objectives did not incorporate the constraints necessary for attacking NILM models.
Motivated by the employed Jacobian property, we propose a novel Jacobian-based method to tackle adversarial NILM problems. We subsequently detail this method in two steps.
\subsubsection{Problem reformulation}
We follow the first-order Taylor expansion~\cite{bolte2021conservative} to perform an approximation:
\begin{equation}\label{eq:taylor}
  f^i(X_{t,w} + E^i_{t,w}) \simeq f^i(X_{t,w}) + J(X_{t,w})E^i_{t,w},
\end{equation}
where $J(X_{t,w}) \in \mathbb{R}^{w*w}$ is the Jacobian matrix of the network at $X_{t,w}$. The Jacobian matrix $J(X_{t,w})$, on the other hand, is a matrix consisting of all partial derivatives of the output of the model $f^i(X_{t,w})$ with respect to the input $X_{t,w}$. It describes the rate of change of the model output with respect to each feature in the input and can be obtained by classical back-propagation approaches. We then make a second approximation $Y^i_{t,w} \simeq f^i(X_{t,w})$ that is the model output approximately equal to the truth data. Based on the two approximations, the \emph{adversarial NILM problem} is changed to find $\tilde{E}^i_{t,w}\in \mathbb{R}^w$:
\begin{equation}\label{eq:adv_new}
\begin{split}
  \tilde{E}^i_{t,w} = \mathop{\arg\max}\limits_{E^i_{t,w}\in C^i_\delta}\parallel J(X_{t,w})E^i_{t,w}\parallel_1\\
  s.t.   C^i_\delta = \{E^i_{t,w} \in \mathbb{R}^w | \parallel E^i_{t,w} \parallel_{\infty}\leq \delta \}.&
\end{split}
\end{equation}
Here, we make a variable change $E^{i'}_{t,w} = {\delta}^{-1}E^i_{t,w}$ and we can further change the \emph{adversarial NILM problem} to (we omit $t, w, and i$ hereafter for simplicity):
\begin{equation}\label{eq:adv_final}
  \tilde{E'} = \mathop{\arg\max}\limits_{E'\in B_{\infty}}\parallel J(X)E'\parallel_1,
\end{equation}
where $B_{\infty}$ is the closed $l_{\infty}$ ball centered at 0 with unit radius. This problem can be further regarded to find a vector $\tilde{E'}$ to attain $\parallel J(X) \parallel_{\infty, 1}$:
\begin{equation}\label{eq:sub_norm}
  \parallel J(X) \parallel_{\infty, 1} =  \mathop{\sup}\limits_{E'\in\mathbb{R}^w\setminus{0}} \frac{\parallel J(X)E'\parallel_1}{\parallel E' \parallel_\infty}
\end{equation}
Assume $J(X)=[J_1, J_2, \ldots, J_w]^T$ ($J_i=[j_{i1}, j_{i2}, \ldots, j_{iw}]$) and $E'=[e'_1, e'_2, \ldots, e'_w]$, such that:
\begin{equation}\label{eq:sub_norm1}
\begin{split}
  \parallel &J(X) \parallel_{\infty, 1} =  \mathop{\sup} \frac{\parallel J(X)E'\parallel_1}{\parallel E' \parallel_\infty}\\
  &= \mathop{\sup}\frac{\parallel[J_1 E', J_2 E', \ldots,J_w E']^T\parallel_1}{\parallel E'\parallel_{\infty}}\\
  &= \mathop{\sup}\frac{|J_1 E'|+\cdots+|J_w E'|}{\mathop{\max}\limits_{i} |e'_i|}\\
  &=\mathop{\sup} \frac{|j_{11}e'_1+\cdots+j_{1w}e'_w|+\cdots+|j_{w1}e'_1+\cdots+j_{ww}e'_w|}{\mathop{\max}\limits_{i} |e'_i|}.
\end{split}
\end{equation}
\begin{algorithm}[t]
    \SetKwInOut{Input}{Input}
    \SetKwInOut{Output}{Output}
    \SetAlgoLined
    \DontPrintSemicolon

    \Input{Matrix $J$, number of iterations $Num$, learning rate $lr$}
    \Output{Best vector $E^*$}
    $w \leftarrow$ Dimension of vector $E'$\;
    $r^* \leftarrow -\infty$, $E^* \leftarrow \text{None}$\;
    Initialize vector $E'$ with $w$ dimensional random values in the range $[-1, 1]$\;
    Initialize Adam optimizer, with learning rate $lr$ and set the first and second moment estimates of the gradient as $m\leftarrow 0$, $v\leftarrow 0$.

    \For{$i$ \text{in range} $Num$}{
        Compute absolute inner products: $\text{abs\_inner\_products} \leftarrow |J \cdot E'|$\;
        Compute sum of absolute inner products: $s \leftarrow \sum \text{abs\_inner\_products}$\;
        Compute maximum absolute element of $E'$: $m \leftarrow \max |E'|$\;
        Compute ratio: $r \leftarrow \frac{s}{m}$\;
        Compute loss: $\text{loss} \leftarrow mean(r)$\;
        Compute gradients of loss respect to $E'$: G =$\frac{\partial loss}{\partial E'}$\;
        \tcp{Update $E'$ using Adam optimizer:}
        Update biased first moment estimate: $m = \beta1 * m + (1-\beta1)*G$\;
        Update biased second raw moment estimate:$v = \beta2 * v + (1-\beta2) * (G)^2$\;
        \tcp{$\beta1$ and $\beta2$ are hyperparameters, typically set to 0.9 and 0.999}
        Compute bias-corrected first moment estimate:$\hat{m}=m / (1 - (\beta1)^i)$\;
        Compute bias-corrected second raw moment estimate:$\hat{v}=v / (1 - (\beta2)^i)$\;
        Update E': $E' = E' + lr * \hat{m} / (\sqrt{\hat{v}} + \varepsilon)$\;
        \tcp{$\varepsilon$ is a small constant for numerical stability, typically set to $1e-8$}
        \If{ $r > r^*$}{
            $r^* \leftarrow r$ \;
            $E^* \leftarrow \text{copy}(E')$\;
        }
    }
    \Return $E^*$\;
    \BlankLine

    \caption{Jaco-Adam: Find Best Vector $E^*$}
    \label{alg-1}
\end{algorithm}

\subsubsection{Problem solution}
We have now reformulated the problem, as expressed in Eq.~(\ref{eq:sub_norm1}).
For both white-box and black-box settings, the jacobian matrix $J(X)$ can be acquired by computing the partial deviates of Seq2seq output $f(X)$ respect to the test sample $X$.

Hence the obtained reformulated problem in Eq.~(\ref{eq:sub_norm1}) becomes an optimization problem.
To search for the optimal $E'$ vector, we can use multiple rounds of random search. However, acquiring an exact optimal solution may be challenging due to the presence of randomness. As the process of randomly generating the vector $E'$ is stochastic, each round may yield different results. Analyzing the equation, we find that it can be interpreted as: the initial perturbation vector $E$ is moved a sufficiently small step $B_{\infty}$ in a certain direction to attain $E'$, such that $E'$ in the neighborhood of $E$ can maximize the objective function  $\frac{\parallel J(X)E'\parallel_1}{\parallel E' \parallel_\infty}$. If the objective is denoted as $loss = \frac{\parallel J(X)E'\parallel_1}{\parallel E' \parallel_\infty}$, the step is intuitively chosen to align with the positive gradient of $loss$ with respect to $E$: $\frac{\partial loss}{\partial E}$.

Specifically, we propose a novel optimization algorithm called \emph{jaco-Adam} to obtain the optimal solution. This algorithm utilizes the Adam optimizer to iteratively find $E'$ that maximizes the ratio presented in Eq.~(\ref{eq:sub_norm1}).
The algorithm is shown in Algorithm~\ref{alg-1}. The gradient $\frac{\partial loss}{\partial E}$ in line $12$ of Algorithm~\ref{alg-1} can be computed using classical back-propagation approaches. We use the $backward()$ function in Pytorch to compute the gradients. Once we obtain the optimal vector $\tilde{E}'$, we can generate the perturbed signal of the smart meter by adding the perturbation ($\delta \tilde{E}'$) to the original aggregated power signal:
\begin{equation}
\begin{split}
   \tilde{X}_{t,w} & =X_{t,w}+\delta\tilde{E}^{i'}_{t,w} \\
     & = (x_{t-\lfloor\frac{w}{2}\rfloor}+\delta\tilde{e}^{i'}_{t-\lfloor\frac{w}{2}\rfloor}, \ldots, x_{t+\lceil\frac{w}{2}\rceil-1}+\delta\tilde{e}^{i'}_{t+\lceil\frac{w}{2}\rceil-1}).
\end{split}
\end{equation}
\subsubsection{Optimization with Practical Consideration}
Algorithm~\ref{alg-1} provides a solution to address the \emph{adversarial NILM problem}. However, it does not factor in the practical scenario that includes an additional constraint, specifically Eq.~(\ref{eq:pra}). In this section, we optimize the solution to satisfy all the constraints in the \emph{practical adversarial NILM problem}.
Our approach is to gradually eliminate the accumulated perturbation.
Specifically, in each iteration, we compute the sum of vector $E'$ and introduce $-\sum E'$ as an additional term to the total loss (thus, $loss$ in line 11 of Algorithm~\ref{alg-1} is modified as $loss \leftarrow mean(r) - \sum E'$) to update $E'$. Intuitively, the subsequent update steps (Line 12 to line 17) for $E'$ in Algorithm~\ref{alg-1} can iteratively maximize loss, consequently reducing the sum of added perturbation $\sum E'$. Additionally, we modify each element of $E'$ by subtracting the mean of all elements of $E'$ to ensure that the sum of $E'$ equals zero in each iteration. This implies that we add the step $E'\leftarrow E' - mean(E')$ after line 17 of Algorithm~\ref{alg-1}. Here, $mean(E') = \frac{\sum_{i=1}^{n} e'_i}{w}$, where $E'=[e'_1, e'_2, ..., e'_w]$. This process does not introduce significant computational overhead. Specifically, the perturbation computation in Algorithm~\ref{alg-1} has a time complexity of $O(Num \cdot batch\_size \cdot w)$, where $Num$ represents the number of iterations and $w$ denotes the length of the perturbation vector. By adjusting the $batch\_size$ and $w$, we can ensure that the perturbation computation is completed in milliseconds, making it feasible for practical deployment.
\section{Experiments}\label{experiment}
We conduct extensive experiments to evaluate our ADV-NILM scheme, focusing on its effectiveness in generating perturbed power signals that significantly degrade the performance of a specific NILM model in both white-box and black-box settings. Additionally, we assess the practicality of our ADV-NILM for supporting accurate billing calculations.
\subsection{Setup}
\subsubsection{Datasets} We evaluate our ADV-NILM using two classical household datasets: REDD~\cite{kolter2011redd} and UK-DALE~\cite{kelly2015uk}.

REDD includes power measurements from 6 US households over about 4 months. The aggregated and individual appliances' power signals are recorded every 1 and 3 seconds, respectively. After removing some abnormal and unusable records, we choose the data from houses 1, 2, and 3 for assessment.
UK-DALE comprises power signals from 6 UK households over a period of 26 months. The aggregated and per-appliance power signals are both recorded every 6 seconds. For the UK-DALE dataset, we select data from houses 1 and 2 after excluding unusable data.

Following previous studies on NILM, we select four main appliances to evaluate the proposed algorithms. Specifically, we set the data from houses 2 and 3 as the training set for REDD, and the data from house 1 as the test set. For UK-DALE, we change the setting to use data from house 1 for training and data from house 2 for testing.
\subsubsection{NILM models}
As stated in Section~\ref{sec:model-des}, for white-box setting we use Seq2seq~\cite{zhang2018sequence}, a CNN-based NILM model, as our target model to generate perturbed signals.
We assume that the architecture and all parameters of the target model are known to the authorized users for the perturbation-generating process.
To evaluate in the black-box setting, we consider four state-of-the-art NILM models: SGN~\cite{shin2019subtask}, MSDC~\cite{He2023msdc}, LDwA~\cite{piccialli2021improving}, and InFocus~\cite{he2023infocus} are target models and their architectures and parameters are unknown for the user. The user applies the perturbation generated from Seq2seq to attack these four target models. These four models are selected as they represent typical solutions spanning across CNN, RNN, and attention mechanisms.
\begin{table*}[htb]
\renewcommand\arraystretch{1.1}
\setlength\tabcolsep{2.5pt}
\caption{The original output VS. the perturbed output of Seq2seq.}
\begin{center}
\scalebox{1.0}{
\begin{tabular}{ c c| c c c c c || c c c c c}
\hline
& \multicolumn{6}{c}{\textbf{Original output}} & \multicolumn{5}{c}{\textbf{Perturbed output}} \\
\hline
Datasets & Metrics& Dishwasher & Fridge & Washing & Microwave & Ave & Dishwasher & Fridge & Washing & Microwave & Ave \\
\hline
\multirow{3}{*}{REDD}   & MAE &19.84&37.36&12.37&27.01&\textbf{24.15}&27.99&52.93&30.31 & 588.44&\textbf{174.91~\baup{150.41}} \\
                        & RSME&72.69&60.33&108.98&153.96&\textbf{98.99}&155.48&99.55&92.2 & 597.99&\textbf{236.30~\baup{137.31}} \\
& SAE &2.81\%&19.08\%&1.97\%&26.35\%&\textbf{12.55\%}&90.20\%&85.23\%&58.24\% & 2547.00\%&\textbf{695.00\%~\baup{682.45\%}} \\
&CORR&0.89 &0.74 &0.92 &0.38 &\textbf{0.74} &-0.01 &0.04 &0.95 & 0.10&\textbf{0.27~\badown{0.47}} \\
\hline
\multirow{3}{*}{UK-DALE} &MAE&22.49&25.13&9.60&8.54&\textbf{16.44}&58.28&32.79&427.31 &31.04 &\textbf{137.35~\baup{120.91}} \\                        &RSME&114.34&40.24&91.72&89.52&\textbf{83.95}&309.34&50.93&462.38 &94.89 &\textbf{229.38~\baup{145.43}} \\
&SAE&0.37\%&26.97\%&72.70\%&58.04\%&\textbf{39.52\%}&79.07\%&55.74\% &4395.12\% &231.90\%&\textbf{1190.45\%~\baup{1150.93\%}} \\
&CORR&0.92 &0.51 &0.67 &0.46 &\textbf{0.64} &-0.13 &0.00 &0.27 &0.36 &\textbf{0.12~\badown{0.52}} \\
\hline
\end{tabular}}
\label{tab:effec}
\end{center}
\end{table*}
\subsubsection{Metrics}
We adopt four widely used metrics in the NILM domain~\cite{shin2019subtask,He2023msdc,he2023infocus,piccialli2021improving} as performance indicators: (1) Mean Absolute Error (MAE), (2) Root Mean Squared Error (RMSE), (3) Normalized Signal Aggregate Error (SAE), and (4) the Pearson Correlation Coefficient (CORR).

\begin{equation}\label{eq:mae}
  MAE = \frac{1}{T}\sum_{t=1}^{T}|\hat{y}^i_t - y^i_t|
\end{equation}
\begin{equation}\label{eq:rsme}
  RSME = \sqrt{\frac{1}{T} \sum_{t=1}^{T} (\hat{y}^i_t - y^i_t)^2}
\end{equation}
\begin{equation}\label{eq:sae}
  SAE = \frac{|\sum_{i=1}^{T}\hat{y}^i_t-\sum_{i=1}^{T}y^i_t|}{\sum_{i=1}^{T}y^i_t}
\end{equation}
\begin{equation}\label{eq:corr}
  CORR = \frac{\sum_{t=1}^{T} (y_t - \bar{y})(\hat{y}_t - \bar{\hat{y}})}{\sqrt{\sum_{t=1}^{T} (y_t - \bar{y})^2 \sum_{t=1}^{T} (\hat{y}_t - \bar{\hat{y}})^2}}.
\end{equation}
In Eqs.~\ref{eq:mae}-\ref{eq:corr}, $\hat{y}^i_t$ represents the estimated power signal given by the NILM model at time step $t$ and $y^i_t$ represents the truth power signal. $y=\{y_1, ..., y_T\}$, $\hat{y} = \{\hat{y}_1, ..., \hat{y}_T\}$, and $\bar{y}$ and $\bar{\hat{y}}$ denote the means of $y$ and $\hat{y}$, respectively.

Higher values of MAE, RSME, and SAE indicate larger errors within the predictions. Conversely, a smaller absolute value of CORR signifies greater disparity between the ground truth and prediction.
\textbf{The primary objective of our proposed approach is thus to produce larger MAE, RSME, and SAE values, and a smaller absolute value of CORR.}
This would highlight a significant deviation between the model's output and the actual ground truth.

\subsubsection{Parameter settings}\label{sec:parameter}
All the involved NILM models are implemented using Python 3.6 and Pytorch 1.12.0+cu116 and trained on machines equipped with GTX 3090 (40G). The batch size for training is set to 32. The Adam optimizer utilized in Algorithm~\ref{alg-1} is configured with a learning rate $lr=0.1$, and the algorithm is executed for $Num$ iterations varying from $1$ to $10$. The Jacobian matrix $J$ of the NILM model is computed using the $backward()$ function in Pytorch. The parameter $\delta$, which multiplies with the obtained vectors from Algorithm~\ref{alg-1}, is adjusted using multiple values (0.02, 0.05, 0.1, 0.15, 0.2). In alignment with previous research on NILM solutions, we establish the length of the input and output window, $w$, as 400 for the REDD, and 200 for the UK-DALE, respectively. These lengths approximately are equate to durations spanning several dozens of minutes.

\subsection{Effectiveness in White-box Setting}\label{sec:effective}
We evaluate ADV-NILM in a white-box setting through the following three experiments: 1) Assessing whether the added perturbation effectively deviates the target NILM model; 2) Comparing ADV-NILM with other Differential Privacy (DP) mechanisms; 3) Comparing ADV-NILM with other adversarial attack strategies.

\textbf{1) With and without our perturbation.} We select Seq2seq as the target model for perturbation generation and then input the perturbed power signal into Seq2seq to observe the resulting changes in the output signals. For evaluation, we focus on data from various appliances, namely, fridge, dishwasher, washing machine, and microwave. The corresponding results are presented in Table~\ref{tab:effec}.

The term ``original output" denotes the input to Seq2seq, specifically, the aggregated power signal, which corresponds to the raw data recorded by the smart meter. Conversely, ``perturbed output" refers to the input of Seq2seq generated via our ADV-NILM model. As demonstrated in Table~\ref{tab:effec}, the perturbed power signals generated by our ADV-NILM model markedly enhance the prediction error of Seq2seq.
For the REDD dataset, the average results for MAE(Watt), RMSE(Watt), and SAE were altered from 24.15, 98.99, and 12.55\% to 174.91, 236.30, and 695.00\%, respectively, reflecting increases of 150.41, 236.30, and 682.45\%. Concurrently, the average CORR decreased from 0.74 to 0.27, revealing a reduction of 0.47. In a similar pattern, for the UK-DALE dataset, the MAE, RSME, and SAE results were modified from 16.44, 83.95, and 39.52\% to 137.35, 229.38, and 1190.45\%, respectively, constituting increases of 120.91, 145.43, and 1150.93\%. The average CORR decreased from 0.64 to 0.12, indicating a reduction of 0.52. \textbf{These results validate the effectiveness of our ADV-NILM model in generating perturbations capable of misleading the target NILM model and yielding inaccurate power consumption data for appliances.} Thus, our ADV-NILM model effectively thwarts adversaries from extracting users' appliance-level consumption patterns.

\begin{figure}[t]
  \centering
  \subfigure[Original output VS. Truth  signal]{\includegraphics[width=1.7in]{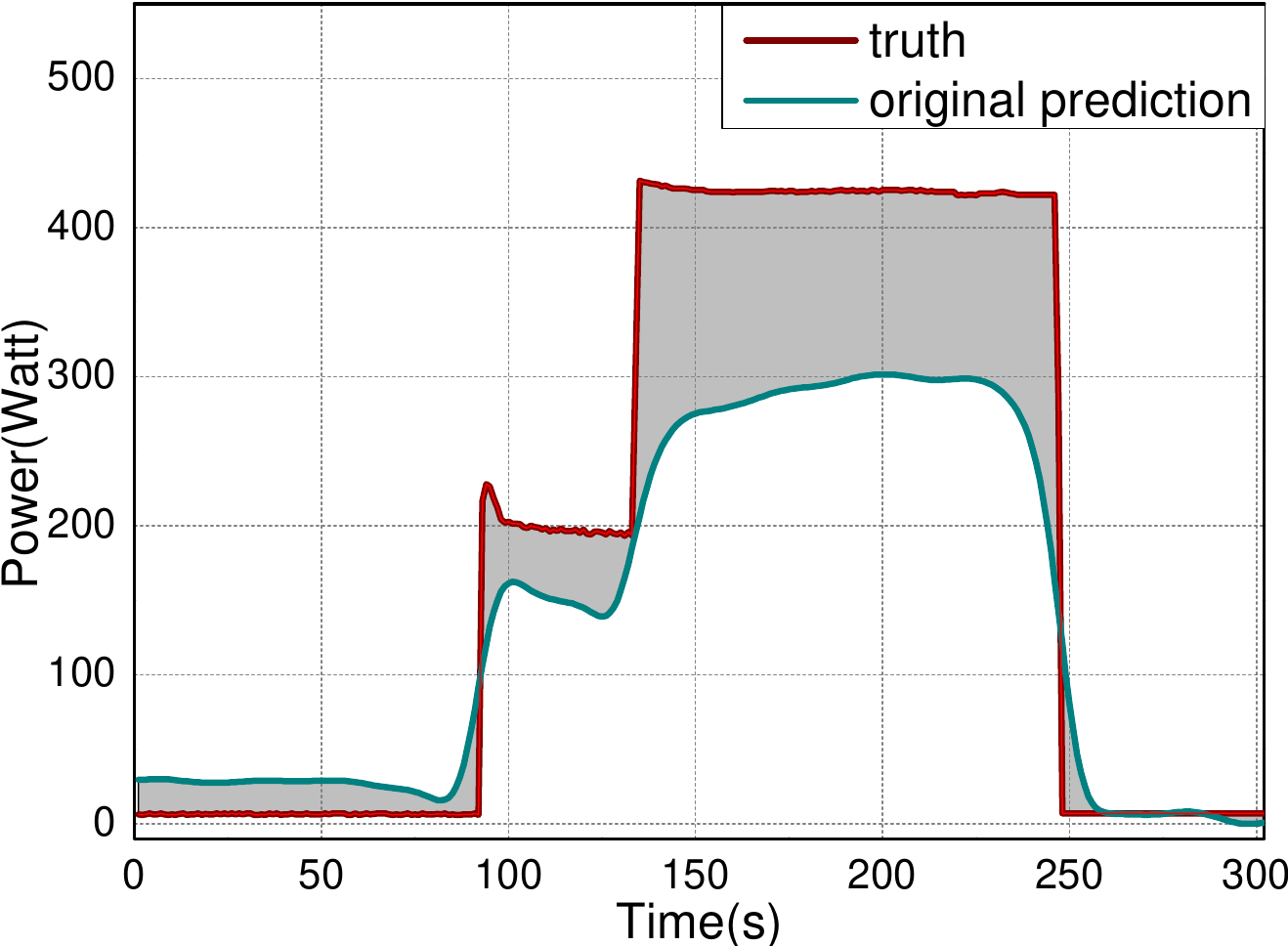} 
  }
  \quad
  \hspace{-0.25in}
  \subfigure[Perturbed output VS. Truth signal]{\includegraphics[width=1.7in]{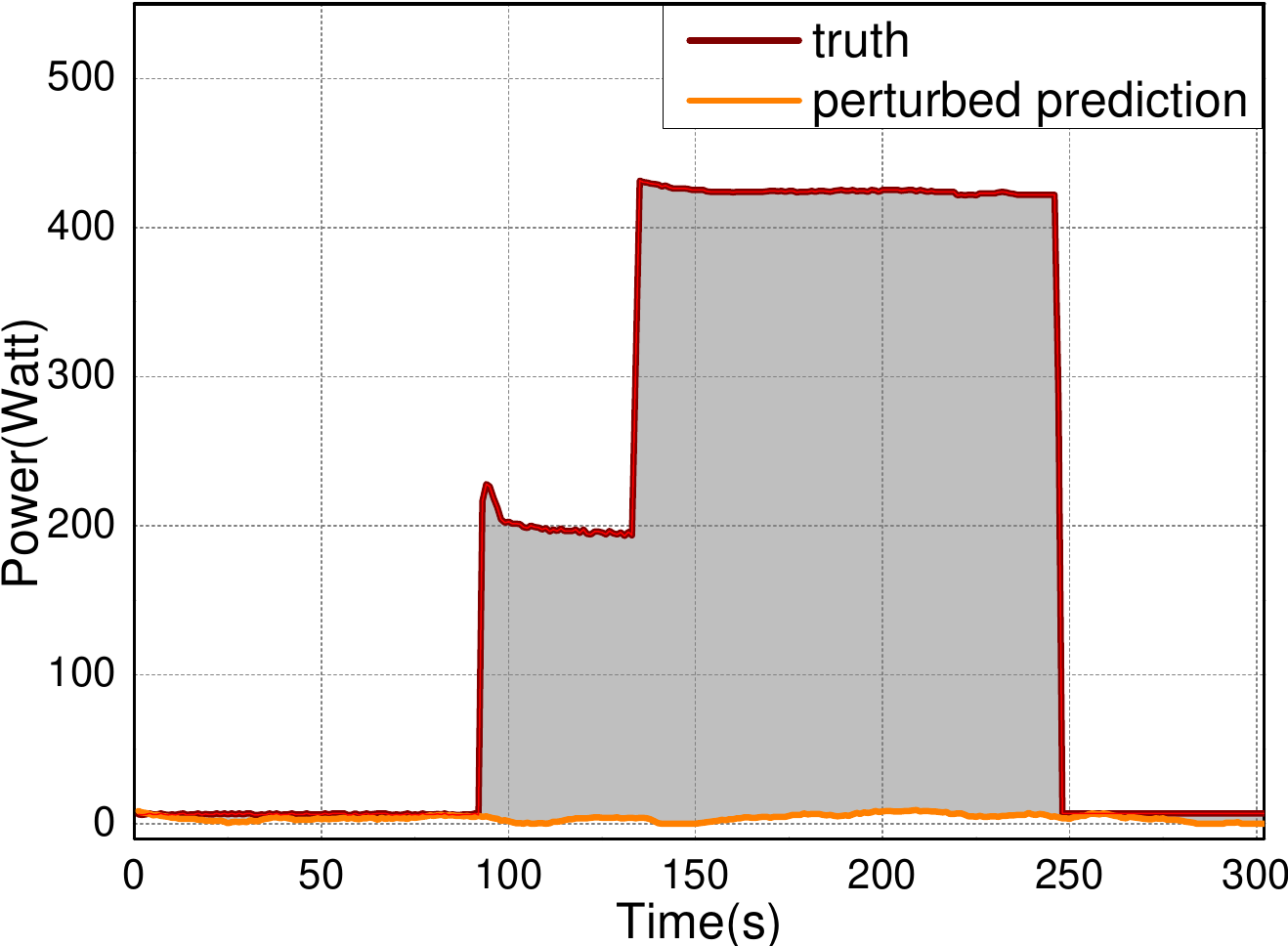}
  }
  \hspace{-0.8in}
    \subfigure[Original output VS. Truth signal]{\includegraphics[width=1.7in]{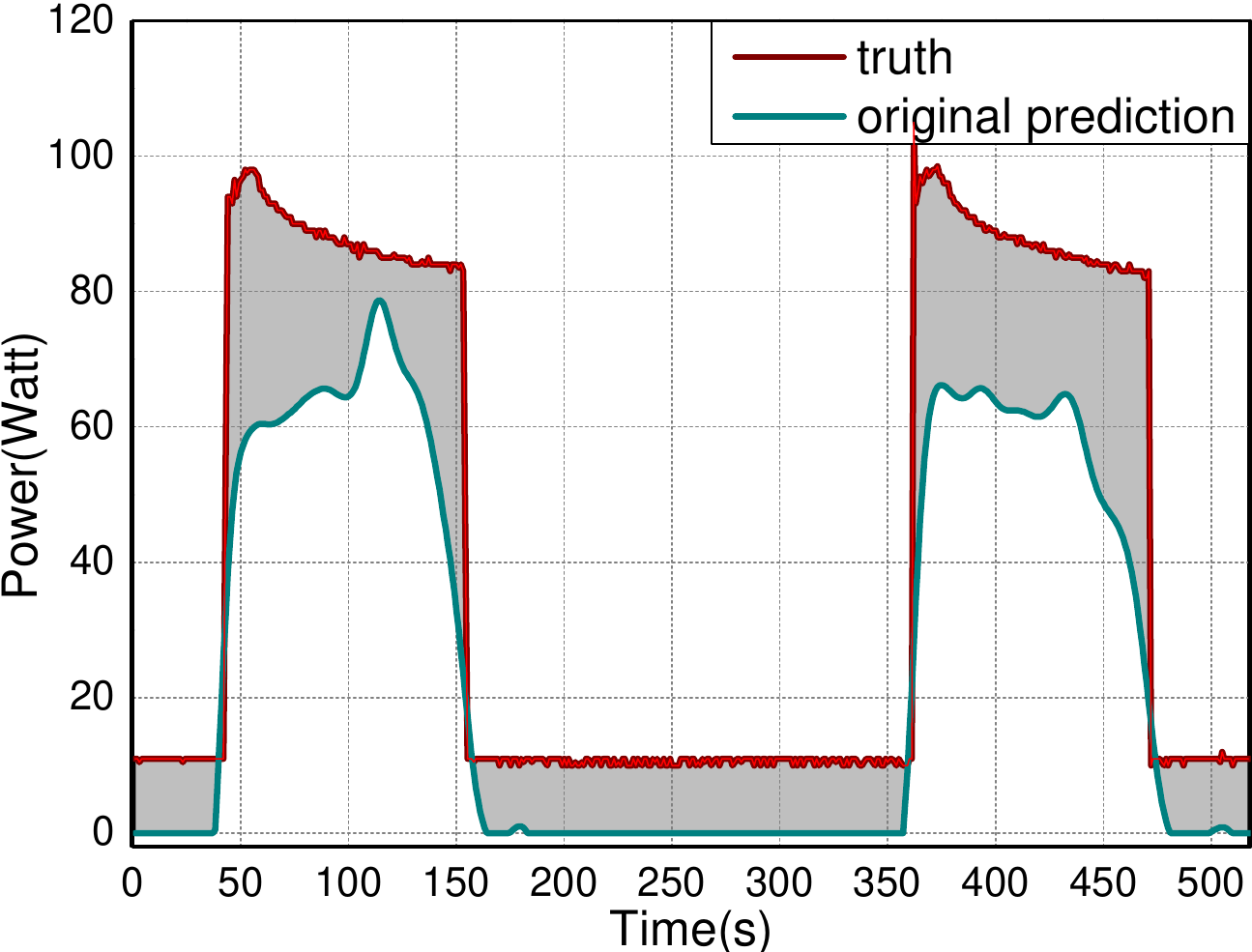}
  }
  \quad
  \hspace{-0.25in}
  \subfigure[Perturbed output VS. Truth signal]{\includegraphics[width=1.7in]{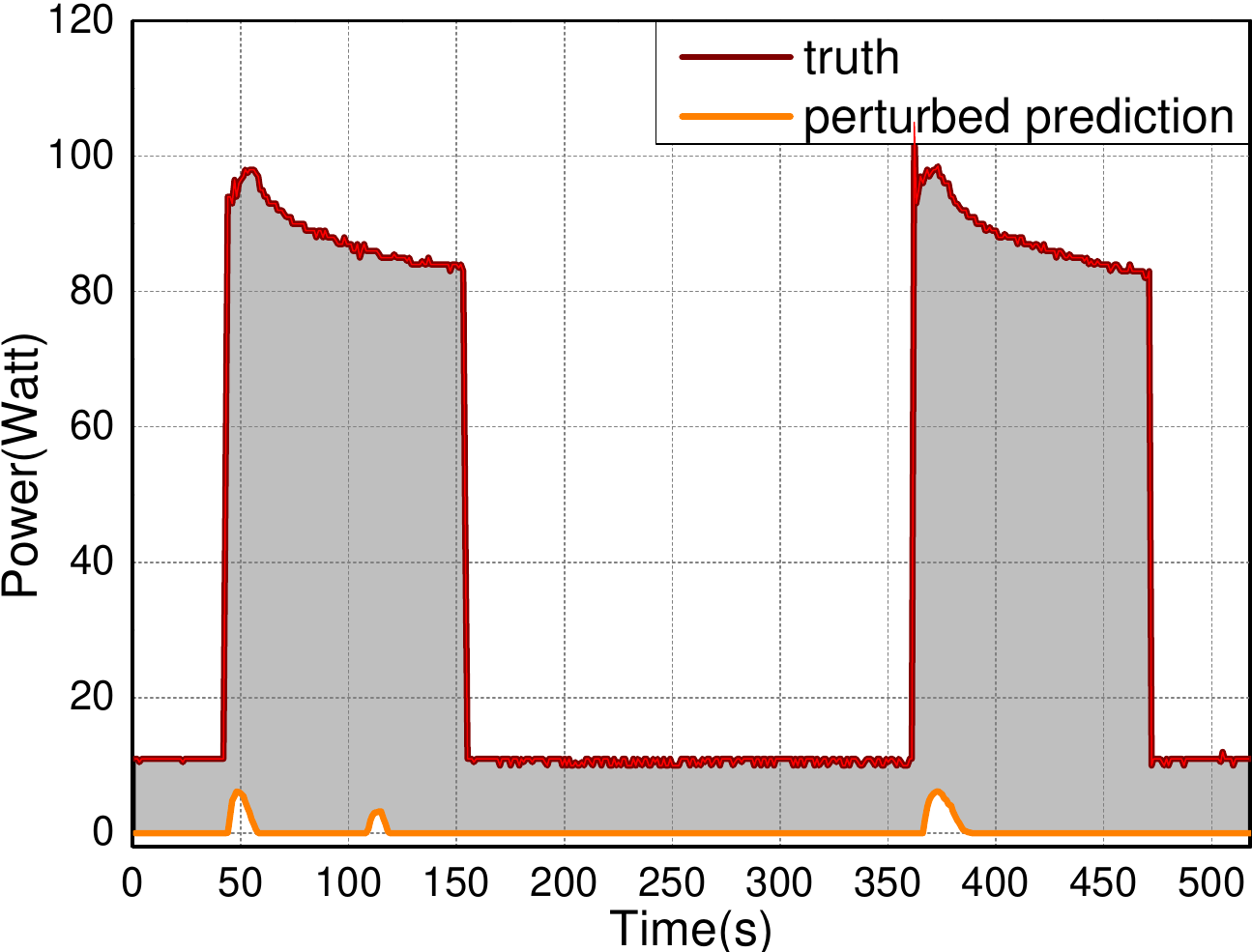}
  }
  \caption{The comparisons between the truth power signal and predicted output from the target NILM model for the fridge. }\label{fig:fridge}
  \vspace{-3mm}
\end{figure}

We further visualize the results to demonstrate the difference between the original output and the perturbed output of Seq2seq. Fig.~\ref{fig:fridge} presents the predicted power signals of appliances in one working cycle, focusing on the fridge for clarity. The first row displays the power curves of the fridge in REDD, while the second row shows the corresponding data for the fridge in UK-DALE.
\begin{table}[htb]
\renewcommand\arraystretch{1.1}
\setlength\tabcolsep{2.3pt}
\caption{Results for the washing machine in the UK-DALE dataset under different $\delta$ settings.}
\begin{center}
\scalebox{1.0}{
\begin{tabular}{ c |c c c c c c c}
\hline
& \multicolumn{6}{c}{\textbf{$\delta$}} \\
\hline
Metrics & 0& 0.02 &0.05 &0.1 & 0.15 & 0.2\\
\hline
MAE & 9.60& 25.02 &60.85 &427.31 &649.10  &802.24 \\
RSME & 91.72 & 128.49 &207.7 &462.38 &662.05  &809.81 \\
SAE & 72.70\% & 128.54\% &541.00\% &4395.12\% & 6695.94\% &8284.85 \%\\
CORR &0.67& 0.37 &0.29 &0.27 &0.28  &0.27 \\
\hline
\end{tabular}}
\label{tab:delta}
\end{center}
\end{table}
In each row of the figure, a comparison is presented between the ground truth power signal and the output signal obtained from Seq2seq. The shaded area between the two curves represents the prediction error. Specifically, the left column displays the original predicted signal produced by Seq2seq, while the right column exhibits the perturbed output. It is evident from the visual representation that the application of our ADV-NILM results in a significant deviation of the predicted output from Seq2seq compared to the ground truth power signal, as evidenced by the larger shaded area between the curves.\textbf{ Furthermore, the on and off patterns for the fridge, discernible in the ground truth, cannot be derived from the perturbed power signal.} This underscores the effectiveness of our ADV-NILM in preventing the disclosure of power consumption patterns.
The visual findings presented in Fig.~\ref{fig:fridge} are consistent with the numerical results shown in Table~\ref{tab:effec}.
\begin{table}[htbp]
\setlength\tabcolsep{6pt}
	\caption{Attacking results for appliances in UK-DALE from the DP mechanism and our ADV-NILM.}
    \begin{center}
	\begin{tabular}{ c c c c c}
		\hline
		Appliances&Metrics&Original&DPDR~\cite{hassan2022differentially}&Ours\\
		\hline
        \multirow{7}{*}[10pt]{Dishwasher}
        &MAE& 22.49&25.58& \textbf{58.28}\\
        &RMSE& 114.34&119.77& \textbf{309.34}\\
        &SAE& 0.37\%&7.52\%& \textbf{79.07\%}\\
        &CORR&0.92 &0.92& \textbf{-0.13}\\
        \hline
        \multirow{7}{*}[10pt]{Fridge}
        &MAE& 25.13&27.78& \textbf{32.79}\\
        &RMSE& 40.24&43.45& \textbf{50.93}\\
        &SAE& 26.97\%&45.81\%& \textbf{55.74\%}\\
        &CORR&0.51 &0.47& \textbf{0.00}\\
       \hline
        \multirow{7}{*}[10pt]{Washing}
        &MAE& 9.60&17.33& \textbf{427.31}\\
        &RMSE& 91.72&96.95& \textbf{462.38}\\
        &SAE& 72.70\%&31.53\%& \textbf{4395.12\%}\\
        &CORR&0.67 &0.49& \textbf{0.27}\\
        \hline
        \multirow{7}{*}[10pt]{Microwave}
        &MAE& 8.54&9.53& \textbf{31.04}\\
        &RMSE& 89.52&88.85& \textbf{94.89}\\
        &SAE& 58.04\%&42.86\%& \textbf{231.90\%}\\
        &CORR&0.46 &0.47& \textbf{0.36}\\
        \hline
	\end{tabular}
    \end{center}
    \label{tab:random}
\end{table}

\begin{table}[htb]
\renewcommand\arraystretch{1.1}
\setlength\tabcolsep{1.5pt}
\caption{Attacking results for the Microwave in REDD and Washing machine in UK-DALE.}
\begin{center}
\scalebox{1.0}{
\begin{tabular}{ l| c c c c|c c c c}
\hline
& \multicolumn{4}{c}{Microwave }& \multicolumn{4}{c}{Washing}\\
\hline
Schemes& MAE & RSME & SAE & CORR &MAE & RSME & SAE & CORR\\
\hline
\emph{Origin} &\emph{27.01} &\emph{153.96}& \emph{26.35\%}& \emph{0.38}&\emph{9.60} &\emph{91.72}& \emph{72.70\%}& \emph{0.67}\\
FGSM &24.95&136.24&16.03\%&0.58&28.15 &140.83 &188.03\% &0.34 \\
PGD&82.68&191.17&243.07\%&0.23 &93.12 &232.60 &918.85\% &0.30\\
GO- &21.34&115.61 &74.51\%&0.39 &56.67 &159.04 &118.64\% &0.31\\
GO+ &84.00&191.94&248.82\%&0.23&93.16&232.20&933.79\%&0.36\\
\textbf{Ours} & \textbf{588.44}& \textbf{597.99}&\textbf{2547.00\%} &\textbf{0.10}&\textbf{427.31} &\textbf{462.38} &\textbf{4395.12\%} &\textbf{0.27}\\
\hline
\end{tabular}}
\label{tab:comparison}
\end{center}
\end{table}

We set $\delta$ in Algorithm~\ref{alg-1} to $0.1$ and obtain the results presented above. The parameter $\delta$ determines the magnitude of the added perturbation from our method. Larger values of $\delta$ correspond to larger noise signals being added. To showcase the effectiveness for different $\delta$ settings, we focus on the washing machine appliance in the UK-DALE dataset as an example. Table~\ref{tab:delta} displays the results, where $\delta=0$ indicates that no perturbation is added to the input signal of Seq2seq, resulting in the original predicted power signal given by Seq2seq.
From Table~\ref{tab:delta}, it is clear that as $\delta$ increases, the performance of the target NILM model deteriorates, with larger MAE, RMSE, and SAE metrics, and smaller the absolute value of CORR. The MAE, RSME, and SAE metrics exhibit an approximate doubling of growth when $\delta$ is set as $0.02$, corresponding to the addition of noise of a relatively small magnitude. Simultaneously, the CORR metric shifts from 0.67 to 0.37, signifying an approximate halving in decrease.

\textbf{2) Our method VS. DP mechanism.}
We have also conducted experiments to highlight the distinction between perturbations of our ADV-NILM and the DP mechanism DRDP~\cite{hassan2022differentially}. In particular, we modified our attack by removing the gradient computation steps and instead generating perturbations using the Laplace noise provided in DPDR. Following the approach outlined in~\cite{hassan2022differentially}, we set the privacy budget to 0.01. It is important to note that the privacy level, or the magnitude of noise, increases as the privacy budget decreases in the DP mechanism. The study~\cite{hassan2022differentially} presents results for privacy budgets set to 0.01, 0.05, 0.1, 0.5, 1, and 2, indicating that the 0.01 privacy budget used in our work represents the highest privacy level achievable with DPDR.

Table~\ref{tab:random} shows the performance of the Seq2Seq model on four appliances from the UK-DALE dataset. Lower values for MAE, RMSE, and SAE, and a higher value for CORR, indicate better prediction accuracy. The results demonstrate that perturbations generated by Algorithm~\ref{alg-1}, which incorporate adversarial attack, significantly degrade the accuracy of NILM model predictions, causing larger errors compared to the ground truth. In contrast, DPDR fails to consistently deceive the NILM model; for appliances such as the washing machine and microwave, the metrics like SAE and CORR even improve compared to the original predictions. Furthermore, when perturbations from DPDR do lead to adversarial effects, the changes in the metrics are much smaller than those observed with our generated perturbations. \textbf{In summary, although the DP mechanism is theoretically proven to achieve privacy by concealing the actual smart meter readings, it cannot effectively deceive NILM models in the same way as our ADV-NILM approach.}

\textbf{3) Our method VS. other adversarial attacks.} We further investigate the effectiveness of our ADV-NILM comparing with other adversarial attack approaches. As there is no existing adversarial attacks aiming at NILM models, we adapt existing attacks such as FGSM~\cite{goodfellow2014explaining}, PGD~\cite{madry2017towards}, and GO~\cite{kong2023adversarial} for attacking NILM models. FGSM and PGD are two classical attacks for image classifiers by exploiting the gradient of target DNN. GO represents a gradient-based attack strategy targeting regressors with a single-time output. It encompasses the GO+ and GO- techniques, correlating respectively with gradient ascent and descent. We use the Seq2seq as the target model to evaluate our ADV-NILM and all the baselines, where the perturbation magnitude is set as $\delta = 0.1$. We take the microwave in REDD and washing machine in UK-DALE as examples and display the attacking results in Table~\ref{tab:comparison}.
``Origin" in Table~\ref{tab:comparison} corresponds to the original results from Se2seq without any attacks. As shown in the table, besides the FGSM and GO- methods, most attacking methods can induce certain deviations in the appliances' signal prediction, attaining an increase in MAE, RSME, and SAE metrics and a decrease in the CORR value. \textbf{Our ADV-NILM method delivers the most effective results for both microwave and washing machine appliances.} For example, our ADV-NILM in capable of obtaining an order of magnitude increase for the MAE, RSME, and SAE metrics, which transition from 9.60, 91.72 and 72.70\% to 427.31, 462.38, and 4395.12\% respectively.
\begin{table*}[htb]
\renewcommand\arraystretch{1.1}
\setlength\tabcolsep{2pt}
\caption{The original output VS. the perturbed output.}
\begin{center}
\scalebox{0.95}{
\begin{tabular}{ p{1.2cm} c| c c |c c |c c |c c |c c}
\hline
\multicolumn{2}{c}{}& \multicolumn{2}{c}{\textbf{Seq2seq\cite{zhang2018sequence}}}& \multicolumn{2}{c}{\textbf{SGN\cite{shin2019subtask}}}&
\multicolumn{2}{c}{\textbf{LDwA\cite{piccialli2021improving}}}&
\multicolumn{2}{c}{\textbf{MSDC\cite{He2023msdc}}}&
\multicolumn{2}{c}{\textbf{InFocus\cite{he2023infocus}}}\\
\hline
Datasets & Metrics& \textbf{original} & \textbf{perturbed} & \textbf{original} & \textbf{perturbed} & \textbf{original} & \textbf{perturbed} & \textbf{original} & \textbf{perturbed} & \textbf{original} & \textbf{perturbed }\\
\hline
\multirow{3}{*}{REDD}   & MAE &37.36&56.33~\baup{18.97}&31.64&64.55~\baup{32.91}&29.24&71.42~\baup{42.18}&31.65&80.61~\baup{48.96}&30.52&71.36~\baup{40.84} \\                        &RSME&60.33&100.57~\baup{40.24}&54.46&97.60~\baup{43.14}&54.28&93.76~\baup{39.48}&54.21&100.31~\baup{46.1}&52.72&95.73~\baup{43.01} \\
& SAE &19.08\%&84.11\%~\baup{65.03\%}&15.47\%&44.68\%~\baup{29.21\%}&12.70\%&12.50\%~\baup{-0.2\%}&12.34\%  &33.33\%~\baup{20.99}&13.88\% &12.80\%~\baup{-1.08\%} \\
&CORR & 0.74&0.05~\badown{0.69} &0.79 &-0.05~\badown{0.84} &0.79 &-0.08~\badown{0.87} &0.78 &0.03~\badown{0.75}&0.80 &-0.04~\badown{0.84} \\
\hline
\multirow{3}{*}{UK-DALE} &MAE&25.13&32.77~\baup{7.64}&23.01&27.13~\baup{4.12}&25.59&28.49~\baup{2.9}&16.10&26.07~\baup{9.97}&15.57&27.09~\baup{11.52} \\                      &RSME&40.24&51.53~\baup{11.29}&39.78&45.42~\baup{5.64}&41.95&44.42~\baup{2.47}&35.19&42.64~\baup{7.45}&34.54&43.91~\baup{9.37} \\
&SAE&26.97\%&59.04\%~\baup{32.07\%}&35.45\%&39.76\%~\baup{4.31\%}&36.24\%&28.26\%~\baup{-7.98\%}&13.20\% &25.85\%~\baup{12.65\%}&10.92\%&32.33\%~\baup{21.41\%} \\
&CORR & 0.51& 0.00~\badown{0.51}&0.56 & 0.34~\badown{0.22}&0.49 &0.27~\badown{0.22} &0.63 &0.36~\badown{0.27}&0.64 &0.33~\badown{0.31} \\
\hline
\end{tabular}}
\label{tab:transfer}
\end{center}
\end{table*}

\begin{table*}[htb]
\renewcommand\arraystretch{1.1}
\setlength\tabcolsep{2.5pt}
\caption{The results for considering the accurate billing operation.}
\begin{center}
\scalebox{1.0}{
\begin{tabular}{ c c| c c c c c || c c c c c}
\hline
& \multicolumn{6}{c}{\textbf{Original output}} & \multicolumn{5}{c}{\textbf{Perturbed output}} \\
\hline
Datasets & Metrics& Dishwasher & Fridge & Washing & Microwave & Ave & Dishwasher & Fridge & Washing & Microwave & Ave \\
\hline
\multirow{3}{*}{REDD}   & MAE &19.84&37.36&12.37&27.01&\textbf{24.15}&25.86&56.33&882.21&35.56&\textbf{249.99~\baup{225.84}} \\                       & RSME&72.69&60.33&108.98&153.96&\textbf{98.99}&153.57&100.57& 894.98&163.65&\textbf{328.19~\baup{229.20}} \\
& SAE &2.81\%&19.08\%&1.97\%&26.35\%&\textbf{12.55\%}&98.52\%  &84.11\% & 2702.21\%&8.00\% &\textbf{723.21\%~\baup{710.66\%}} \\
&CORR& 0.89&0.74 &0.92 &0.38 &\textbf{0.74} &-0.05 &0.05 &0.66 &-0.03 &\textbf{0.15~\badown{0.59}} \\
\hline
\multirow{3}{*}{UK-DALE} &MAE&22.49&25.13&9.60&8.54&\textbf{16.44}&58.38&32.77&420.55 &30.67 &\textbf{135.59~\baup{119.15}} \\
                        &RSME&114.34&40.24&91.72&89.52&\textbf{83.95}&309.33&51.53  &456.13 &94.99 &\textbf{227.99~\baup{144.04}} \\
&SAE&0.37\%&26.97\%&72.70\%&58.04\%&\textbf{39.52\%}&78.87\%&59.04\% &4325.00\% &227.02\%&\textbf{1172.48\%~\baup{1132.96\%}} \\
&CORR&0.92 &0.51 &0.67 &0.46 &\textbf{0.64} &-0.14 &0.00 &0.28 &0.36 &\textbf{0.12~\badown{0.52}} \\
\hline
\end{tabular}}
\label{tab:prac}
\end{center}
\end{table*}
\subsection{Effectiveness in Black-box Setting}\label{sec:transfer}
In Section~\ref{sec:effective}, we highlighted the effectiveness of our ADV-NILM in a white-box setting, where both the target attack model and the perturbation generation model are Seq2seq. In this section, we extend our investigation to examine the transferability of our proposed method to other state-of-the-art NILM models, emulating a black-box setting. Specifically, we generate perturbed signals based on Seq2seq and use these signals to deceive other NILM models, including SGN~\cite{shin2019subtask}, MSDC~\cite{He2023msdc}, LDwA~\cite{piccialli2021improving}, and InFocus~\cite{he2023infocus}.

The specific results are presented in Table~\ref{tab:transfer}, with a focus on the performance evaluation for the fridge. The ``original" in the table represents inputting the original aggregated power measurements to the NILM models to obtain the prediction output, while ``perturbed" refers to inputting the perturbed aggregated power signals to the NILM models to obtain the perturbed output. The table clearly demonstrates that the perturbation generated from the Seq2seq model significantly impairs the predictive performance of other NILM models. Regarding the MAE, RSME, and SAE metrics, barring a few data points that depict a minor decline, all other data segments consistently display a substantial increase.
For instance, considering the MAE metric in the context of the REDD dataset, the performances of the SGN, LDwA, MSDC, and InFocus models deteriorate by $\frac{32.91}{31.64}=104.01\%$, $\frac{42.18}{29.24}=144.25\%$, $\frac{48.96}{31.65}=154.69\%$, and $\frac{40.84}{30.52}=133.90\%$, respectively. Additionally, \textbf{regarding the CORR metric, the perturbed signal leads all SGN, LDwA, MSDC, and InFocus models to exhibit a considerable decrease.}
These findings compellingly illustrate the effective transferability of the perturbation signal generated from the Seq2seq model to other NILM models, thereby demonstrating the transferability of our ADV-NILM.
\begin{figure}[t]
  \centering
  \subfigure[Original output VS. Truth signal]{\includegraphics[width=1.7in]{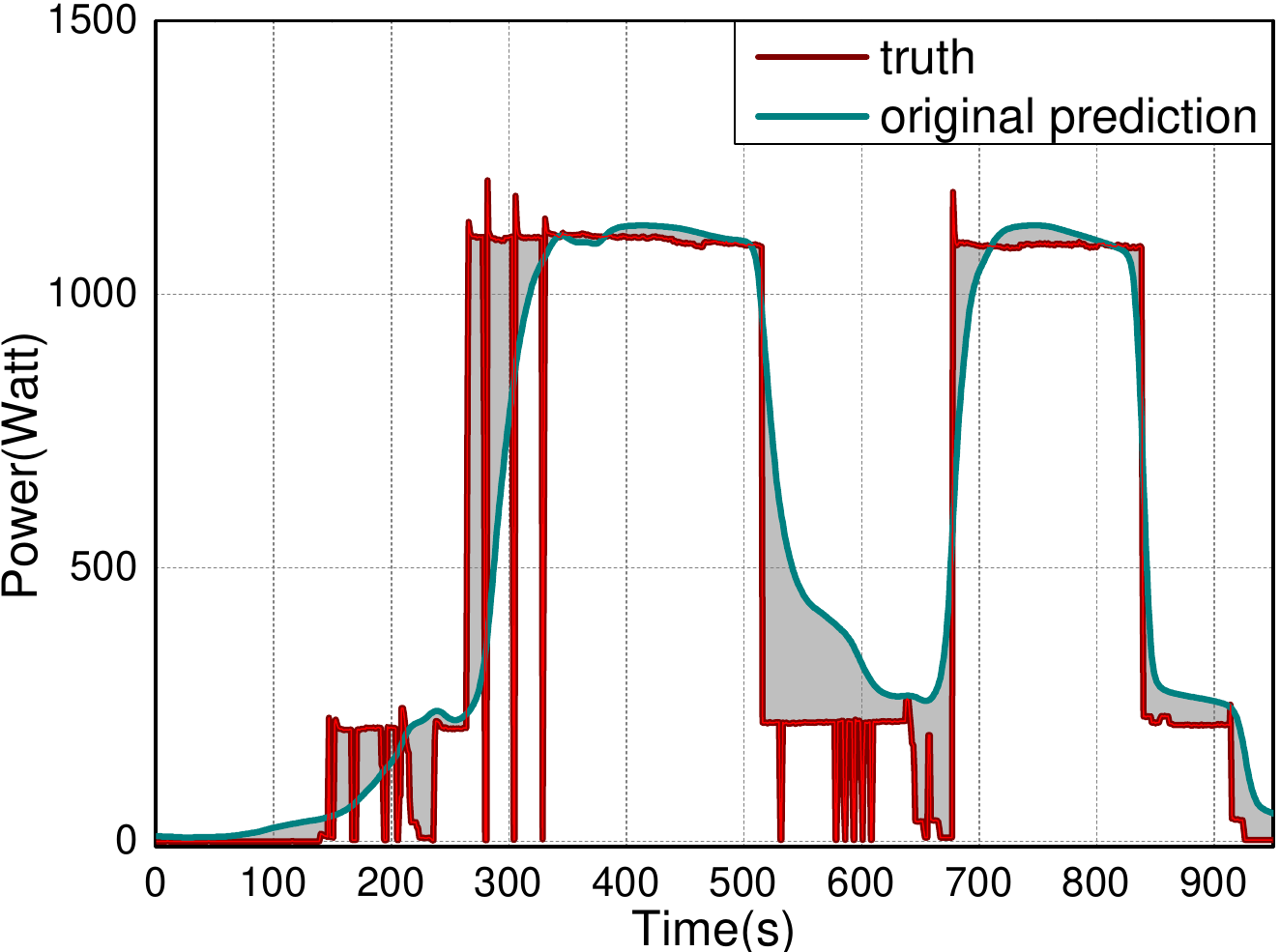}
  }
  \quad
  \hspace{-0.25in}
  \subfigure[Perturbed output VS. Truth signal]{\includegraphics[width=1.7in]{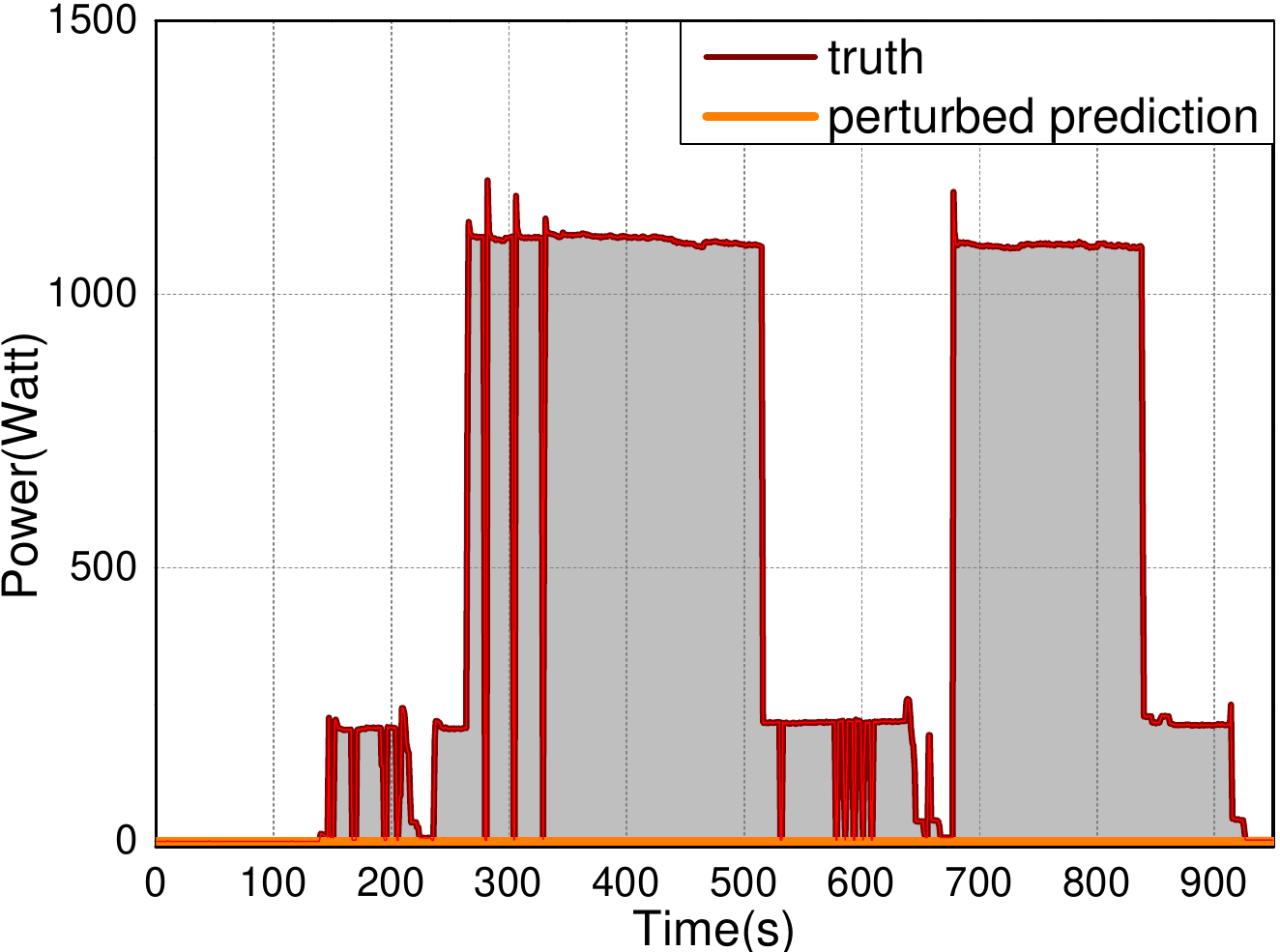}
  }
  \hspace{-0.8in}
    \subfigure[Original output VS. Truth signal]{\includegraphics[width=1.7in]{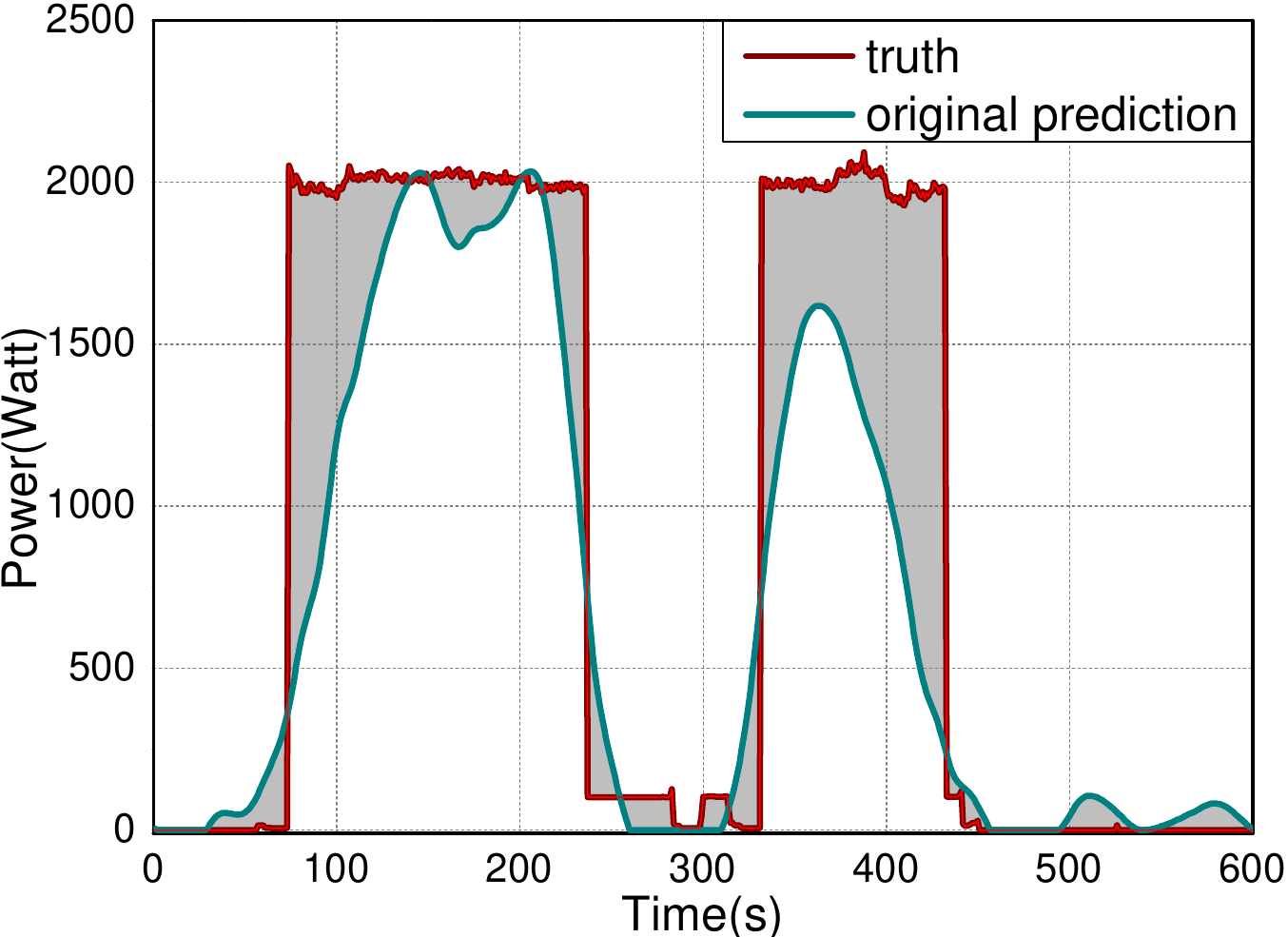}
  }
  \quad
  \hspace{-0.25in}
  \subfigure[Perturbed output VS. Truth signal]{\includegraphics[width=1.7in]{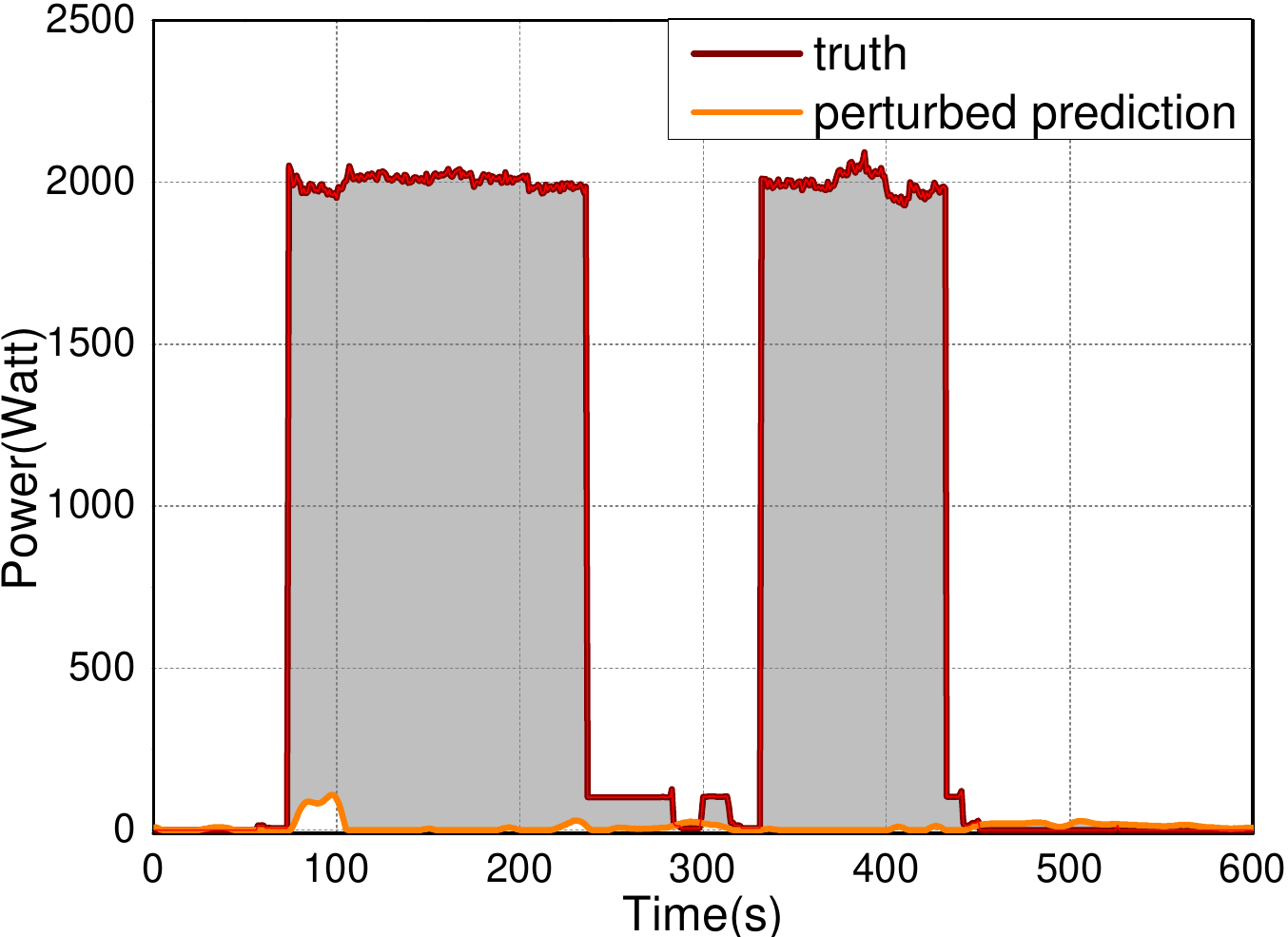}
  }
  \caption{The comparisons between the truth power signal and predicted output from the target NILM model for the dishwasher. (The left column corresponds to the original predicted output and the right column is for the output after the perturbing operation by our ADV-NILM model with practicality consideration.)}\label{fig:practical}
\end{figure}
\begin{table}[htb]
\renewcommand\arraystretch{1.1}
\setlength\tabcolsep{3pt}
\caption{Results for the fridge in the UK-DALE dataset under different $Num$ settings.}
\begin{center}
\scalebox{1.0}{
\begin{tabular}{ c |c c c c c c c}
\hline
& \multicolumn{6}{c}{\textbf{$Num$}} \\
\hline
Metrics &$0$& $1$ &$3$ &$5$ & $6$ & $8$ &$15$\\
\hline
MAE & 22.49& 41.79 &51.12 &58.38 &67.28 &62.41  &49.78 \\
RSME & 114.34 & 230.69 &305.46 &309.33 &310.13 &310.25  &309.87 \\
SAE & 0.37\% & 60.59\% &92.93\% &78.87\% &0.6147 & 71.00\% &100 \%\\
CORR &0.92& 0.87 &0.33 &-0.14 &-0.36 &-0.27   &-1.23 \\
\hline
\end{tabular}}
\label{tab:ite}
\end{center}
\end{table}
\subsection{Practicality}\label{sec:prac}
The findings presented in Section~\ref{sec:effective} and Section~\ref{sec:transfer} have substantiated the merits of our proposed ADV-NILM scheme. However, the perturbation generation algorithm utilized in Algorithm~\ref{alg-1} does not fulfill the requirement for accurate billing calculations. To address this limitation, the \emph{practical adversarial NILM problem} is introduced, which incorporates a constraint ensuring the sum of consecutive perturbations added within a window is zero. This facilitates accurate billing calculations. In this section, we proceed to assess the performance of our ADV-NILM scheme with the incorporation of accurate billing consideration.

Similar to the process in Section~\ref{sec:effective}, we set $\delta$ to be $0.1$ and present the comparisons between the original outputs and perturbed outputs of the target NILM model Seq2seq in Table~\ref{tab:prac}. The results demonstrate a substantial increase in MAE, RSME, and SAE, along with a decrease in the absolute value of CORR, for nearly all appliances in both the REDD and UK-DALE datasets. For example, the MAE of the washing machine in the REDD dataset increases from $12.37$ to $882.21$, representing a nearly hundredfold augmentation. Additionally, the average results for MAE, RSME, and SAE in both datasets also exhibit impressive growths. For REDD, the average MAE, RSME, and SAE changed from $24.15$, $98.99$, and $12.55\%$ to $249.99$, $328.19$, and $723.21\%$ respectively, achieving growths of $225.84$, $229.20$, and $710.66\%$. For UK-DALE, the corresponding results changed from $16.44$, $83.95$, and $39.52\%$ to $135.59$, $227.99$, and $1172.48\%$, representing growths of $119.15$, $144.04$, and $1132.96\%$ respectively. Conversely, the absolute value of CORR experiences a significant reduction. In the REDD dataset, the average absolute CORR value shifts from 0.74 to 0.15, realizing a decrease of 0.59. A comparable outcome is observed in the UK-DALE dataset, where it alters from 0.64 to 0.12, signifying a decrease of 0.52.
\textbf{These findings suggest that the need for accurate billing does not compromise the efficacy of our method in deceiving NILM models.}

Fig.~\ref{fig:practical} is also utilized to visually demonstrate the disparity between the original and the perturbed output obtained from the NILM model. Specifically, Fig.~\ref{fig:practical} showcases the results for the dishwasher appliance. The two rows in the figure correspond to the results from the REDD and UK-DALE datasets, respectively. It is evident from the figure that the predicted power signal from Seq2seq becomes significantly distinct from the actual data after the perturbation from our ADV-NILM is added. As the shadow area between the two curves in the subfigures of the right column is observed to be larger compared to that of the left column. Notably, the `turn-on' pattern is almost entirely eliminated from the perturbed signal. This suggests that the additional constraint for accurate billing calculation does not compromise the efficacy of our ADV-NILM.
Furthermore, as indicated in Section~\ref{sec:parameter}, considering the window length is roughly dozens of minutes, \textbf{our technique can achieve precise billing, even within the context of traditional dynamic pricing strategies.}
%

All previous results were obtained by setting the number of iterations, denoted as $Num$, in the algorithm to $5$. The results of the dishwasher in the UK-DALE dataset are presented in Table~\ref{tab:ite} to illustrate the effectiveness of $Num$. In the table, $Num =0$ corresponds to the original performance of Seq2seq without any perturbation. The table reveals that as the number of iterations increases, the performance of Seq2seq worsens, with larger MAE, RSME, and SAE metrics, and a smaller absolute value of CORR. However, the absolute value of CORR begins to increase after the fifth iteration. Therefore, all previous experiments in our study, including the baseline metrics of PGD, GO+, and GO- used in Section~\ref{sec:effective}, were conducted with $Num =5$.

\section{Conclusions}
In this paper, we have presented a novel scheme aimed at safeguarding the privacy of power-consuming patterns against NILM models. Our investigation begins with formulating the problem of adversarial NILM, and through rigorous theoretical analysis, we have devised a solution algorithm that capitalizes on the Jacobian of the target NILM model to generate perturbations.
To ensure accurate billing for users, we have then enhanced the algorithm by introducing an additional constraint. Empirical validation of the effectiveness of our proposed scheme has been conducted through extensive experiments on real-world power datasets, namely REDD and UK-DALE.
The experimental results evidently demonstrate that our proposed scheme significantly undermines the predictive effectiveness of the targeted NILM model, effectively preserving users' privacy at the appliance level.

\bibliographystyle{IEEEtran}
\bibliography{IEEEexample}
\vspace{-10mm}
\begin{IEEEbiography}[{\includegraphics[width=0.8in,height=1in,clip,keepaspectratio]{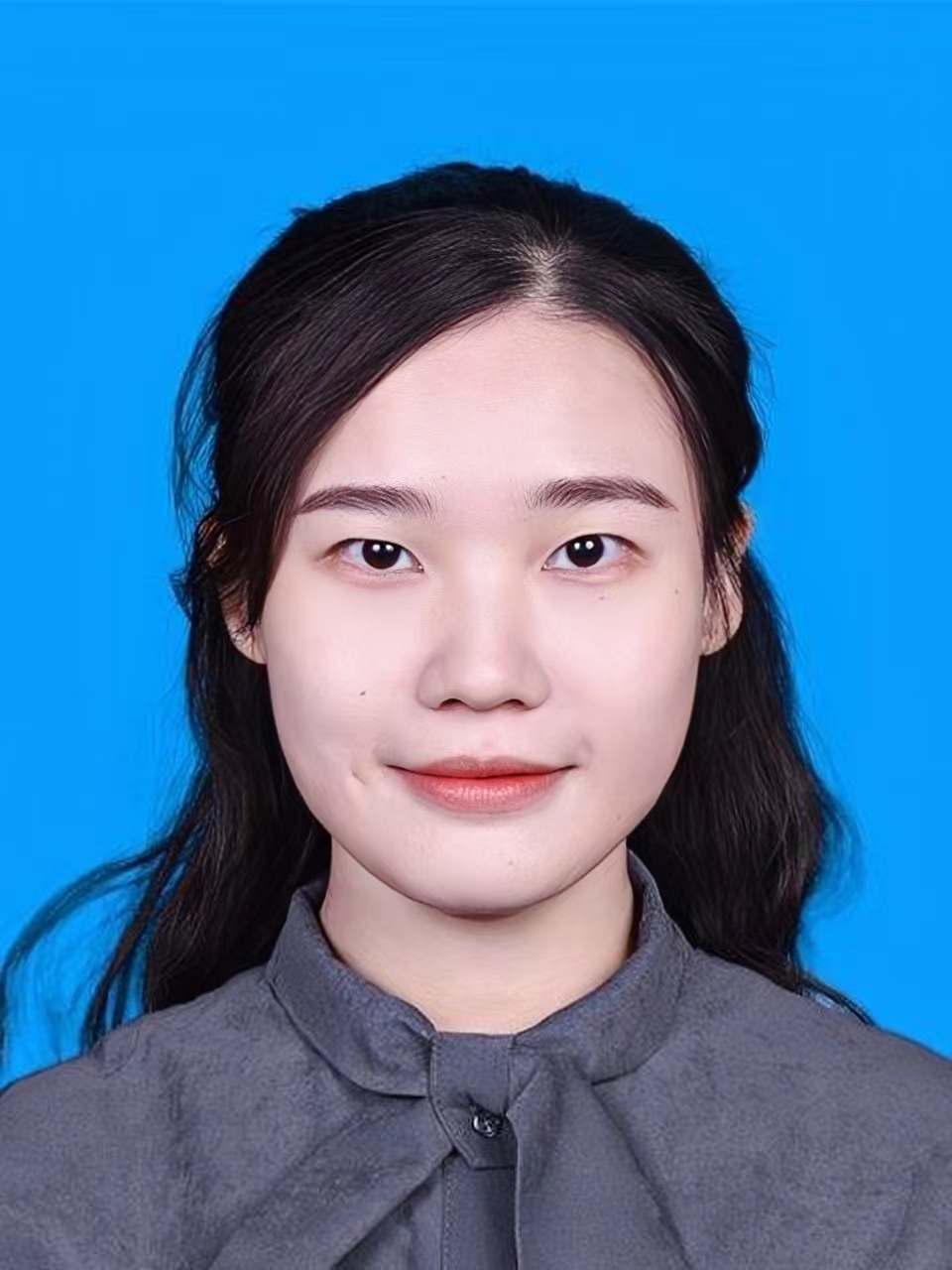}}]{Jialing He}
\noindent
received the M.S. and Ph.D. degrees from the Beijing Institute of Technology, Beijing, China, in 2018 and 2022, respectively, where she is currently an research assistant professor in college of computer science, Chongqing University, Chongqing, China. Her current research interests include differential privacy, user behavior mining, and Blockchain.
\end{IEEEbiography}

\vspace{-10mm}
\begin{IEEEbiography}[{\includegraphics[width=0.8in,height=1in,clip,keepaspectratio]{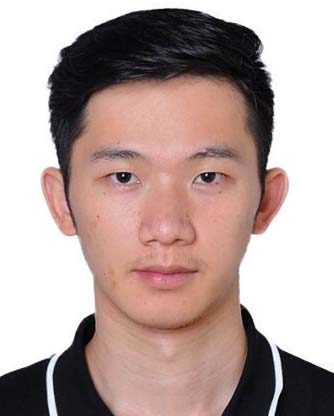}}]{Jiacheng Wang}
\noindent
received the Ph.D. degree from the School of Communication and Information Engineering, Chongqing University of Posts and Telecommunications, Chongqing, China. He is cur-rently a Research Associate in computer science and engineering with Nanyang Technological University, Singapore. His research interests include wireless sensing, semantic communications, and metaverse.
\end{IEEEbiography}

\vspace{-10mm}
\begin{IEEEbiography}[{\includegraphics[width=0.8in,height=1in,clip,keepaspectratio]{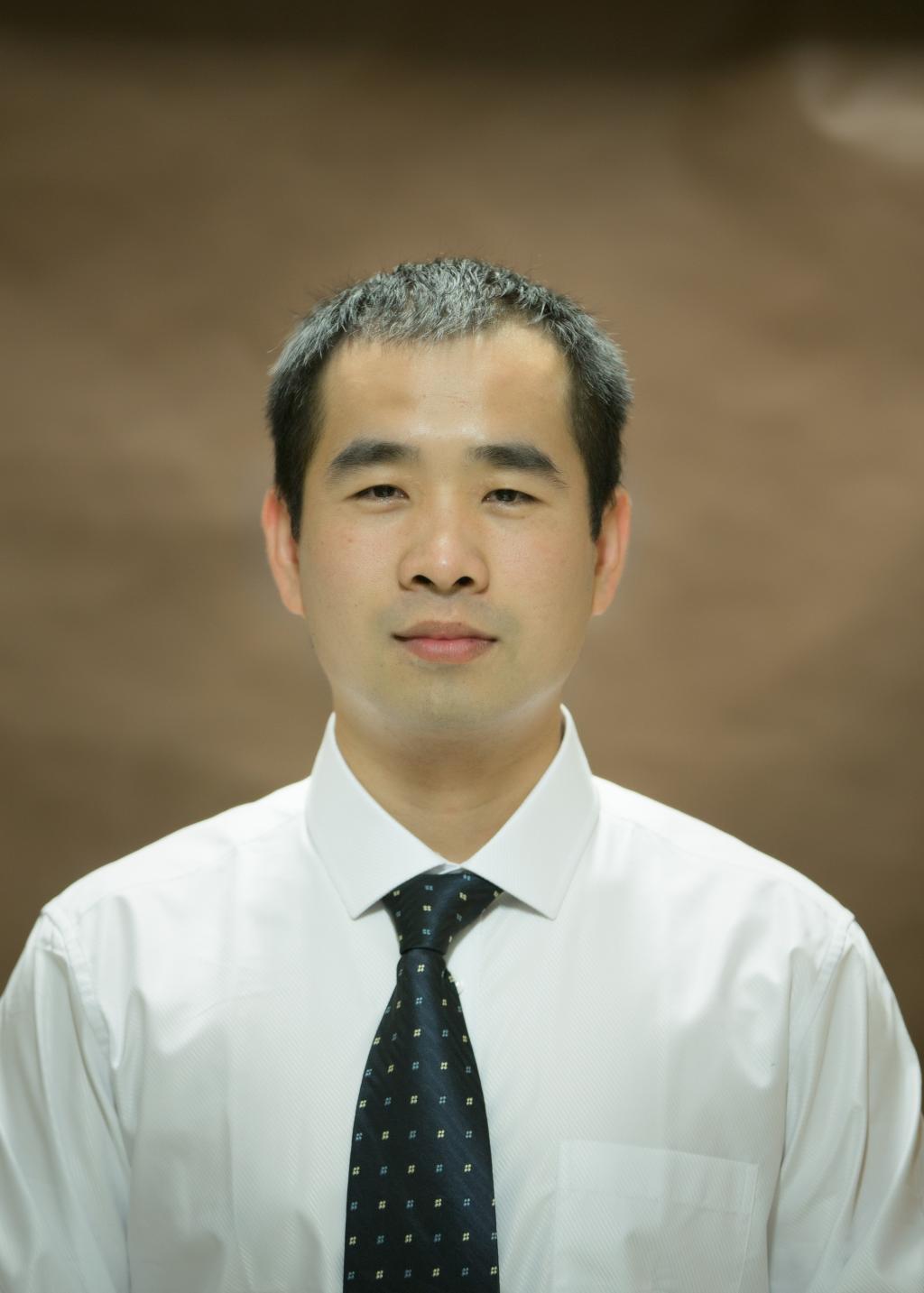}}]{Ning Wang}
\noindent
received the Ph.D. degree in information and communication engineering from the Beijing University of Posts and Telecommunication in 2017. He was an Engineer at Huaxin Post and Telecommunications Consulting Design Company Ltd. from 2012 to 2013. He was with the Department of Electrical and Computer Engineering, George Mason University, as a PostDoctoral Scholar from 2017 to 2020. He is currently a Professor with the College of Computer Science Chongqing University. His current research interests are in physical layer security, machine learning, RF fingerprinting, and cyber-physical systems security and privacy
\end{IEEEbiography}
\vspace{-10mm}
\begin{IEEEbiography}[{\includegraphics[width=1in,height=1.25in,clip,keepaspectratio]{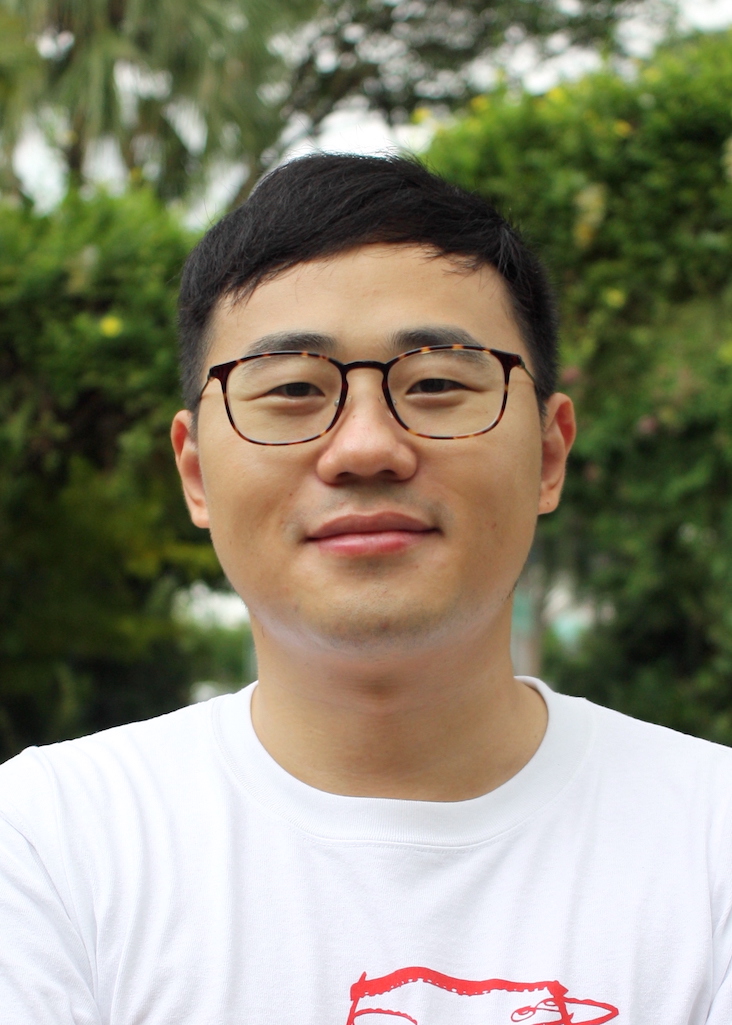}}]%
  {Shangwei Guo}
  is an associate professor in College of Computer Science, Chongqing University. He received the Ph.D. degree in computer science from Chongqing University, Chongqing, China at 2017. He worked as a postdoctoral research fellow at Hong Kong Baptist University and Nanyang Technological University from 2018 to 2020. His research interests include machine learning security and multimedia security.
  \end{IEEEbiography}
\vspace{-10mm}
\begin{IEEEbiography}[{\includegraphics[width=0.8in,height=1in,clip,keepaspectratio]{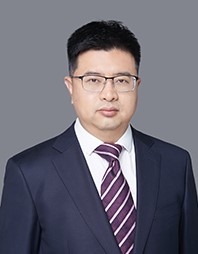}}]{Liehuang Zhu}
\noindent
    is a full professor in the School of Cyberspace Science and Technology, Beijing Institute of Technology. He is selected into the Program for New Century Excellent Talents in University from Ministry of Education, P.R. China. His research interests include Internet of Things, Cloud Computing Security, Internet and Mobile Security.He has published over 100 SCI-indexed research papers in these areas, as well as a book published by Springer. He serves on the editorial boards of three international journals, including IEEE IoT Journal, IEEE Network, and IEEE TVT.He won the Best Paper Award at IEEE/ACM IWQoS 2017 and IEEE TrustCom 2018.
\end{IEEEbiography}
\vspace{-10mm}
\begin{IEEEbiography}[{\includegraphics[width=0.8in,height=1in,clip,keepaspectratio]{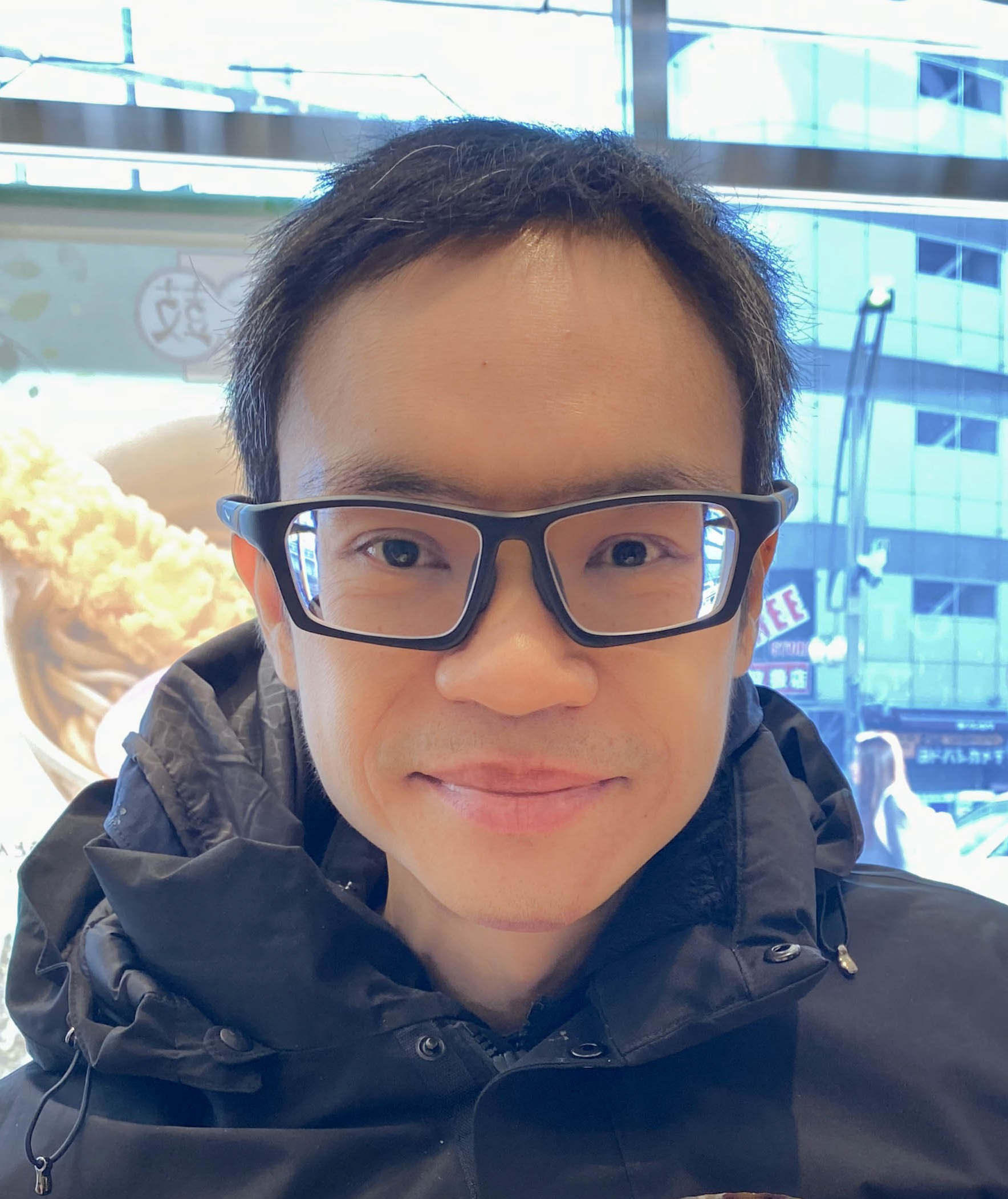}}]{Dusit Niyato}
\noindent
    is a professor in the College of Computing and Data Science, at Nanyang Technological University, Singapore. He received B.Eng. from King Mongkuts Institute of Technology Ladkrabang (KMITL), Thailand and Ph.D. in Elec-trical and Computer Engineering from the University of Manitoba, Canada. His research interests are in the areas of sustainability, edge intelligence, decentralized machine learning, and incentive mechanism design.
\end{IEEEbiography}
\vspace{-10mm}
\begin{IEEEbiography}[{\includegraphics[width=0.8in,height=1in,clip,keepaspectratio]{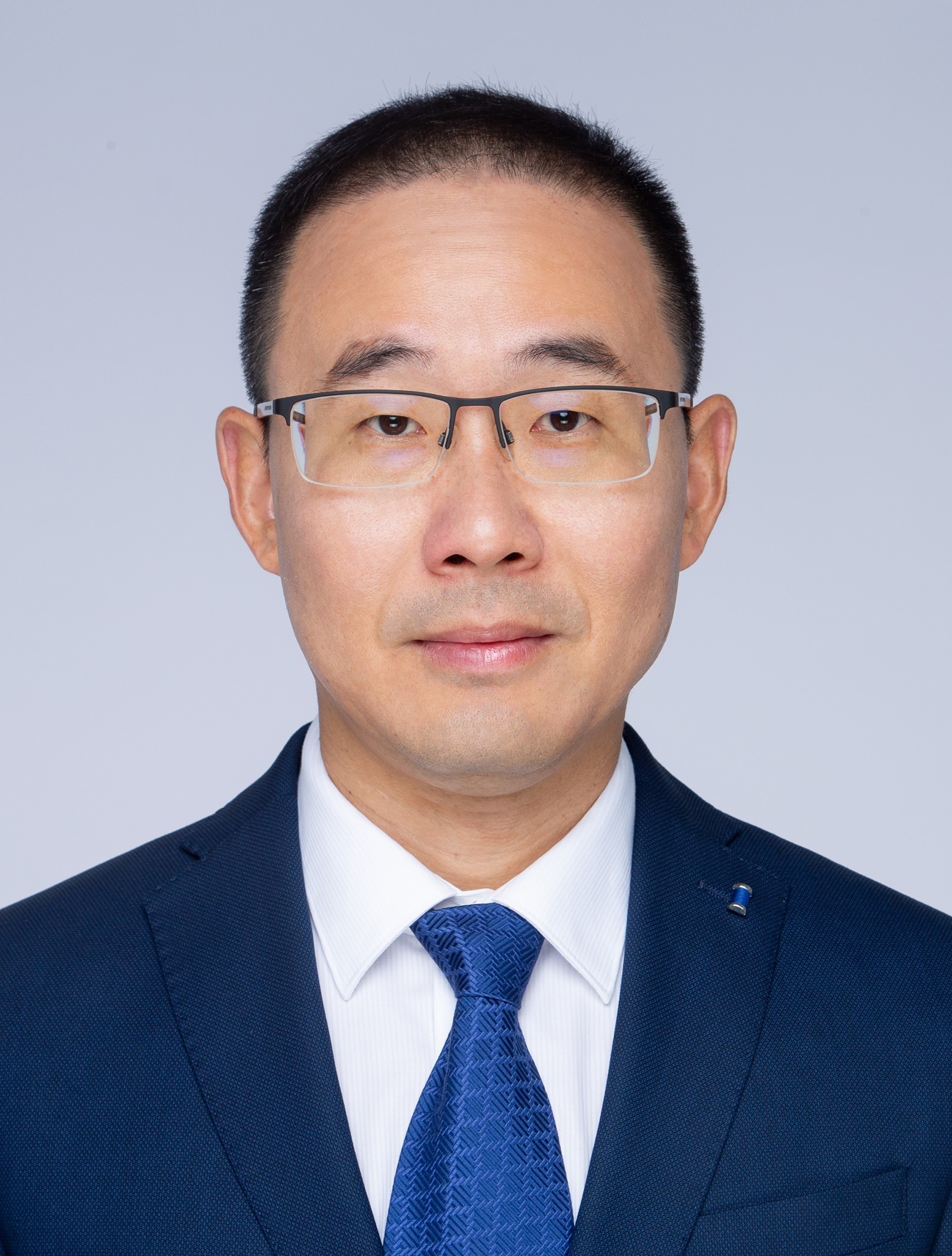}}]%
    {Tao Xiang}
    received the BEng, MS and PhD degrees in computer science from Chongqing University, China, in 2003, 2005, and 2008, respectively. He is currently a Professor of the College of Computer Science at Chongqing University. Prof. Xiang’s research interests include multimedia security, cloud security, data privacy and cryptography. He has published over 100 papers on international journals and conferences. He also served as a referee for numerous international journals and conferences.
\end{IEEEbiography}
\end{document}